\documentclass[journal,twocolumn,10pt]{IEEEtran}
\IEEEoverridecommandlockouts 
\usepackage{epsfig,latexsym}
\usepackage{float}
\usepackage{indentfirst}
\usepackage{amsmath}
\usepackage{amssymb}
\usepackage{times}
\usepackage{subfigure}
\usepackage{pifont}
\usepackage{psfrag}
\usepackage{cite}
\usepackage{url}
\usepackage{lastpage}
\usepackage{bm}
\usepackage{color}
\usepackage{fancyhdr}
\usepackage[center]{caption2}
\usepackage{balance}
\usepackage{fancyhdr,lastpage}
\usepackage{hyperref}
\usepackage{epstopdf}
\usepackage{amsthm}
\newtheorem{algorithm}{Algorithm}
\begin{document}
	\title{Improving Energy Efficiency of Uplink RSMA via Rydberg Atomic Quantum Receiver}
	\author{\IEEEauthorblockN{Xinya Shen,~Zheng Chu,~\emph{Member, IEEE}, 
			David Chieng,~\emph{Senior Member, IEEE}, Chiew Foong Kwong,~\emph{Senior Member, IEEE},~Qihao Peng,~\emph{Member, IEEE}, Zhengyu Zhu,~\emph{Senior Member, IEEE}, Pei Xiao,~\emph{Senior Member, IEEE} and Jianyue Zhu,~\emph{Member, IEEE}
		}
		\thanks{Xinya Shen, Zheng Chu, David Chieng, and Chiew Foong Kwong are with the Department of Electrical and Electronic Engineering, University of Nottingham Ningbo China, Ningbo 315100, China, and also with the Next Generation Internet of Everything Laboratory, University of Nottingham Ningbo China, Ningbo 315100, China (Email: xinya.shen@nottingham.edu.cn; andrew.chuzheng7@gmail.com; David.Chieng@nottingham.edu.cn; Chiew-Foong.Kwong@nottingham.edu.cn)}
		\thanks{Qihao Peng and Pei Xiao are with the 5GIC \& 6GIC, Institute for Communication Systems (ICS), the University of Surrey, Guildford, GU2 7XH, UK. (Email: q.peng@surrey.ac.uk; p.xiao@surrey.ac.uk)}
		\thanks{Zhengyu Zhu is with the School of Electrical and Information Engineering, Zhengzhou University, Zhengzhou 450001, China. (e-mail: zhuzhengyu6@gmail.com)}
        \thanks{Jianyue Zhu is with the College of Electronic and Information Engineering, Nanjing University of Information Science and Technology, Nanjing 210044, China (Email: zhujy@nuist.edu.cn)}
    }
	\vspace{-0.5in}
	\maketitle 
	\thispagestyle{empty}
	\begin{abstract}
		This paper investigates an uplink rate-splitting multiple access (RSMA) network empowered by a Rydberg atomic quantum receiver (RAQR) to maximize energy efficiency (EE). To overcome the amplitude-only measurement limitation of the considered direct readout RAQR, we devise a phase alignment mechanism that linearizes the received signal while retaining full channel energy, thereby enabling tractable optimization. The resulting non-convex fractional EE problem is transformed via Dinkelbach's method. For IoT scenarios with symmetric rate requirements, we derive exact closed-form power allocation from the Karush-Kuhn-Tucker (KKT) conditions, which yields a globally optimal bisection search with negligible computational latency. For asymmetric rate requirements, we develop a generalized block coordinate descent-based power allocation (BCD-PA) framework that eliminates reliance on expensive interior-point solvers. We further propose a low-complexity paired allocation (LC-Paired) that exploits user pairing and closed-form bandwidth solutions via the Lambert-W function. Simulation results confirm that the KKT-based scheme attains the highest EE, while the BCD-PA offers a flexible complexity-performance trade-off. The LC-Paired scheme drastically reduces computational cost with only moderate EE degradation.
        \end{abstract}
	
\section{Introduction}
\label{sec:introduction}

The rapid deployment of massive IoT devices in next-generation (6G) wireless networks has dramatically increased the demands for massive connectivity, high spectral efficiency, and sustainable energy consumption. To address these challenges, RSMA has emerged as a flexible multiple-access paradigm. By splitting user messages into common and private parts, RSMA effectively bridges SDMA and NOMA and enables efficient interference management. Meanwhile, maximizing energy efficiency (EE) remains a critical objective for the green evolution of 6G communication networks.

\subsection{Literature Review}

\subsubsection{Evolution Towards Flexible and Efficient Multiple Access}
To address the spectral-efficiency limitations of orthogonal multiple access (OMA), power-domain NOMA was introduced, allowing multiple users to share the same time-frequency resources through superposition coding and successive interference cancellation (SIC). However, NOMA's rigid SIC ordering and sensitivity to channel uncertainty have motivated the search for more flexible frameworks. RSMA has emerged as a generalized paradigm that bridges OMA, NOMA, and spatial-division multiple access by splitting user messages into common and private parts, thereby enabling robust interference management even without perfect channel state information at the transmitter \cite{mao2022rate, clerckx2021noma}. We focus on uplink RSMA because, in massive IoT, the energy bottleneck lies in battery-limited devices. In contrast, downlink RSMA would require such receivers at the user side, which is impractical for cost- and energy-constrained IoT terminals. This motivates our focus on uplink RSMA empowered by a BS-side RAQR. Early investigations of uplink RSMA confirmed its potential for sum-rate improvement and enhanced user fairness \cite{zhu2017rate, yang2020sum}.

This flexibility has made RSMA effective in a wide range of emerging communication scenarios, including RIS-assisted systems \cite{sun2023joint}, UAV communications \cite{liu2023downlink}, satellite networks \cite{10643592, can2025data}, V2X scenarios \cite{liu2025joint}, and massive IoT applications employing random access protocols \cite{11052254}. Advanced RSMA variants have also been proposed to further improve capacity and spatial efficiency, including distributed uplink RSMA (DU-RSMA) \cite{11161157}, user cooperation enhanced RSMA \cite{9852986}, and pinching-antenna-based RSMA. These developments demonstrate the versatility of RSMA as a unifying multiple access framework for diverse 6G use cases.

A critical challenge specific to uplink RSMA is the optimization of the decoding order at the receiver, whose complexity grows factorially with the number of users, rendering joint power and decoding-order optimization computationally demanding for large-scale deployments \cite{10330667}. Hierarchical approaches to splitting and decoding have been explored to manage this complexity \cite{sadeghabadi2025achievable}. In satellite RSMA systems, channel-dependent power allocation and decoding order selection have been investigated to improve outage performance \cite{10643592}, and data-oriented designs have been analysed under practical constraints \cite{can2025data}. Notably, the work in \cite{10330667} proposed a low-complexity user-pairing strategy that transforms the joint power-and-decoding-order problem into a tractable power-and-bandwidth allocation problem with closed-form solutions. The concept of user pairing, widely examined in NOMA to reduce complexity \cite{ding2015impact}, offers a promising direction for practical RSMA design. Despite these advances, existing RSMA EE optimizations still rely heavily on computationally burdensome iterative algorithms. Consequently, leveraging user-pairing concepts to derive exact closed-form solutions and reduce iterative inner-loop processing remains necessary for latency-sensitive and resource-constrained IoT deployments.

\subsubsection{Rydberg Atomic Quantum Receivers}
The detection of ultra-weak uplink signals from energy-constrained devices constitutes a fundamental physical bottleneck in next-generation networks, as the thermal noise floor and limited linearity of conventional RF receivers impose irreducible sensitivity constraints \cite{degen2017quantum}. Quantum-assisted sensing has emerged as a transformative approach to overcome these classical limitations, and Rydberg atomic quantum receivers (RAQRs) have been identified as a promising candidate for future wireless communication systems \cite{gong2025rydberg, chen2025harnessing}.

The sensing mechanism of RAQRs exploits the exaggerated electric dipole moments of highly excited Rydberg states together with electromagnetically induced transparency (EIT) to transduce incident RF electric fields into optical readouts \cite{saffman2010quantum}. This atomic-level field-to-optical conversion offers several advantages over conventional electronic receivers, including quantum-limited sensitivity, immunity to Johnson--Nyquist thermal noise, ultra-broad spectral tunability, and high linearity \cite{anderson2021atomic, artusio2022modern}. These advantages have been highlighted in surveys on quantum sensing and its prospective role in 6G systems \cite{wang2022quantum, cui2025rydberg}.

Experimental studies have gradually advanced RAQR-based communication from proof-of-concept demonstrations to practical prototypes. Early works verified the feasibility of recovering amplitude-modulated and frequency-modulated waveforms using room-temperature atomic vapour cells \cite{fan2015atom, meyer2018digital, yuan2023rydberg}. A major breakthrough was achieved with the atomic superheterodyne architecture, where a microwave-dressed local oscillator substantially amplifies the atomic response and improves sensitivity by several orders of magnitude over direct EIT readout \cite{jing2020atomic}. Subsequent studies further characterized RAQRs for RF field metrology and sensing \cite{simons2021rydberg, fancher2021rydberg, yuan2023quantum}, evaluated wideband operation \cite{meyer2020assessment}, compared noise temperature with conventional receivers \cite{santamaria2022comparison}, and demonstrated RF phase measurement via atom-based mixing \cite{simons2019rydberg}.

On the theoretical side, equivalent baseband channel models have been developed for both SISO and MIMO RAQR systems, enabling the analysis of achievable performance gains and capacity limits relative to conventional antenna arrays \cite{cui2025towards, gong2025equivalent, liu2025sa}. A defining characteristic of standard RAQR implementations, however, is that they perform magnitude-only measurements of the incident electric-field envelope and discard phase information. This amplitude-only constraint introduces phase ambiguity and prevents the direct application of conventional coherent detection and interference cancellation techniques \cite{wang2020tight}. To address this issue, dedicated signal processing methods have been developed, including RAQR-specific MIMO precoding \cite{cui2024mimo}, hardware-constrained channel estimation under magnitude observations \cite{xu2025channel}, and multi-user uplink reception schemes for the amplitude-only regime \cite{gong2025rydbergMIMO}. The integration of RIS with atomic receivers has also been explored to provide controlled phase references for detection \cite{peng2025ris}. Further extensions include multi-band and multi-carrier operation for spectrum aggregation \cite{cui2025rydberg2, wang2025multi}, as well as spatial signal processing functions such as multi-target direction-of-arrival estimation and polarization-aware detection using a single atomic receiver \cite{gong2025rydberg2, chen2025polarization}. Nevertheless, mapping this fundamentally non-linear, amplitude-only detection mechanism into the multi-stream interference structure of RSMA remains a challenging open problem.

\subsection{Motivation and Contributions}

Although existing research has made significant progress in RSMA for next-generation networks and has separately explored the physical mechanism of RAQR, there remains a clear research gap in their integrated design for communication systems. Specifically, by leveraging the ultra-weak uplink signal detection capability of RAQR, the EE of resource-constrained IoT devices can be improved at the physical layer. This direction remains largely unexplored in existing RSMA research. Moreover, although EE is a key goal of RSMA, most existing strategies focus on power allocation and interference management at the algorithm level. The phase-coherent RAQR array provides a new physical-layer method, which is crucial for energy-constrained 6G IoT applications \cite{santamaria2022comparison}. At the same time, by further combining the user pairing strategy, this framework transforms the joint power and decoding-sequence optimization problem into a lower-complexity form and provides a feasible path for deriving exact closed-form solutions \cite{10330667}. Therefore, it is necessary to study a RAQR-based RSMA framework to systematically address hardware measurement limitations, algorithmic complexity, and system EE in 6G wireless deployment.

To bridge this gap, we conceive a novel uplink RSMA system assisted by an RAQR at the base station and develop a comprehensive optimization framework. To the best of the authors' knowledge, RAQR-empowered RSMA networks remain largely unexplored. The main contributions of this paper are summarized as follows:
\begin{itemize}
    \item We establish a novel equivalent system model for RAQR-empowered uplink RSMA. To overcome the fundamental physical limitation and phase ambiguity of atomic measurements, we design a phase alignment mechanism that linearizes the received signal while retaining full channel energy, thereby enabling tractable optimization.

    \item We formulate a non-convex EE maximization problem and recast it via the Dinkelbach transform. To address heterogeneous scenarios, we devise tailored solutions: closed-form optimal power allocation via KKT conditions for IoT networks with symmetric rate requirements, a generalized block coordinate descent (BCD) framework for asymmetric rate-requirement settings, and a low-complexity paired allocation that exploits strong--weak user pairing to obviate iterative convex solvers.

    \item The proposed frameworks feature high scalability and computational efficiency, enabling globally optimal bisection searches with negligible latency and eliminating reliance on interior-point methods. Simulations confirm that Scheme A achieves the highest EE by fully leveraging multiuser diversity, while Scheme B offers a compelling trade-off, reducing computational cost by orders of magnitude with only moderate EE degradation, making both well suited for RSMA-based 6G deployments.
\end{itemize}

The remainder of this paper is organized as follows. Section~\ref{sec:System Model} presents the system model, including the RAQR-empowered uplink RSMA transmission framework and the achievable rate analysis for both Scheme A and Scheme B. Section~\ref{sec:framework} details the proposed unified three-layer decomposition framework, covering the inner-layer phase alignment via the extended EM-GS algorithm, the middle-layer resource allocation strategies for symmetric and asymmetric rate-requirement scenarios, and the outer-layer Dinkelbach iteration for EE maximization. Simulation results and performance comparisons are provided in Section~\ref{sec:simulation}. Finally, Section~\ref{section:Conclusion} concludes the paper.

\subsection{Notations}
For clarity, the following notation is adopted throughout this paper. Italic letters denote scalars, bold lowercase letters denote vectors, and bold uppercase letters denote matrices. $\mathbb{R}$ and $\mathbb{C}$ represent the real and complex number fields, respectively. The operators $|\cdot|$, $\|\cdot\|$, and $\|\cdot\|_F$ denote the element-wise modulus, the Euclidean norm, and the Frobenius norm, respectively. $\mathcal{R}\{\cdot\}$, $\mathcal{I}\{\cdot\}$, and $\angle(\cdot)$ denote the real part, imaginary part, and phase of a complex variable, respectively. The superscripts $(\cdot)^T$, $(\cdot)^H$, and $(\cdot)^*$ stand for the transpose, Hermitian transpose, and the optimal value under current conditions, respectively. The superscript $(\cdot)^{(i)}$ denotes the value at the $i$-th iteration. Moreover, $\hat{x}$ and $\tilde{x}$ denote the estimated and transformed values of variable $x$, while $(\cdot)^\text{opt}$ refers to the optimally phase-aligned equivalent real-valued channel. $\mathbb{E}[\cdot]$ denotes the expectation operator. $\mathcal{Q}(\cdot)$ represents the pulse amplitude modulation (PAM) demapping operation. Finally, $\mathcal{CN}(\boldsymbol{\mu}, \boldsymbol{\Sigma})$ denotes a circularly symmetric complex Gaussian distribution with mean vector $\boldsymbol{\mu}$ and covariance matrix $\boldsymbol{\Sigma}$. The operator $\circ$ represents the Hadamard product, and $\boldsymbol{I}_M$ denotes an $M \times M$ identity matrix.

\section{System Model and Problem Statement}
\label{sec:System Model}

\subsection{System Model}

\begin{figure}[!htbp]
    \centering
    \includegraphics[width=1\linewidth]{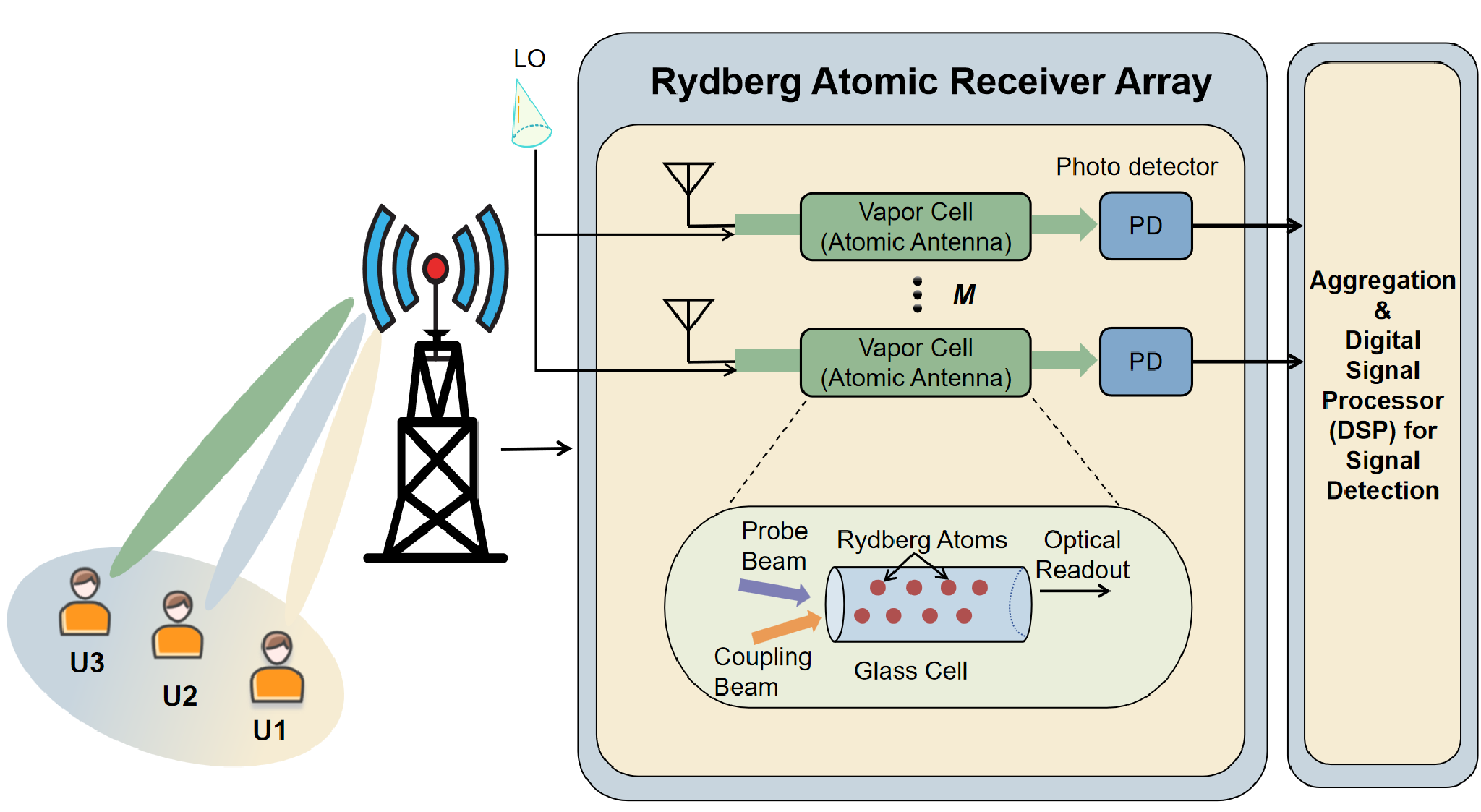}
    \caption{A three-user uplink RSMA network with RAQR.}
    \label{fig:system model}
\end{figure}

This section considers a single-cell uplink network where a single base station (BS) serves three users, denoted as $\mathrm{U}_k$ ($k\in\{1,2,3\}$). While the detailed physical principles of the RAQR can be found in \cite{cui2025towards}, our focus is on the equivalent communication model. As illustrated in Fig.~\ref{fig:system model}, the BS utilizes an $M$-element RAQR instead of a conventional RF front-end. This array comprises $M$ vapor cells functioning as atomic antennas, each filled with Rydberg atoms and coupled to a dedicated photodetector (PD). The optical readouts from these PDs are aggregated and digitally processed for signal detection.

Unlike conventional OMA or NOMA, the adopted RSMA framework \cite{10330667} enables flexible interference management via message splitting. Specifically, each user $\mathrm{U}_k$ splits its information message into a common part and a private part. These parts are encoded into independent Gaussian symbols $x_k^c$ and $x_k^p$, and transmitted with allocated powers $p_k^c$ and $p_k^p$, respectively.

Based on this transmission protocol, the transmitted signal of user $k$ in each symbol interval is expressed as
\begin{align}\label{eq:tx_signal}
s_k = \sqrt{p_k^c}\,x_k^c + \sqrt{p_k^p}\,x_k^p,
\end{align}
where $x_k^c$ and $x_k^p$ are independent real-valued PAM symbols with unit power. The complex baseband signal vector $\mathbf{y}\in\mathbb{C}^{M\times 1}$ at the output of the RAQR is modeled as
\begin{align}
\mathbf{y} = \sum_{k=1}^{K}\mathbf{h}_k s_k + \mathbf{b} + \mathbf{n}, \label{eq:rx-signal}
\end{align}
where \(\mathbf{h}_k \in \mathbb{C}^{M \times 1}\) denotes the equivalent baseband channel vector from user \(\mathrm{U}_k\) to the \(M\)-element atomic array, \(\mathbf{b}\in \mathbb{C}^{M\times 1}\) is the deterministic local-oscillator (LO) reference signal, and \(\mathbf{n}\sim\mathcal{CN}(\mathbf{0},\sigma_n^2\mathbf{I}_M)\) is the additive noise. Equation~\eqref{eq:rx-signal} represents the equivalent complex baseband model derived from the Rydberg EIT principle. Under small-signal conditions, the down-converted optical field is a linear superposition of the users' symbols \(\{s_k\}\) weighted by the channel vectors \(\{\mathbf{h}_k\}\), plus the deterministic LO term \(\mathbf{b}\). Consequently, the PD output follows the linear model in~\eqref{eq:rx-signal}. Although the dominant microscopic noise in practical Rydberg receivers is signal-dependent shot noise \cite{wang2020tight}, this paper adopts the circularly symmetric complex Gaussian approximation based on the central limit theorem to ensure analytical tractability for subsequent rate analysis and resource allocation.

Having established the equivalent baseband model in~\eqref{eq:rx-signal}, we now clarify the physical origin of the channel vector \(\mathbf{h}_k\). It is not an abstract parameter but is explicitly determined by the atomic parameters of the Rydberg vapor cells. For the \(m\)-th cell, let \(\Omega_p^{(m)}\) and \(\Omega_c^{(m)}\) denote the probe and coupling Rabi frequencies, \(\Delta_p^{(m)}\) and \(\Delta_c^{(m)}\) the corresponding detunings, and \(\Gamma_{21},\Gamma_{31},\Gamma_{32}\) the decay rates of the relevant atomic transitions. Under the small-signal condition, the probe-transition density matrix element \(\rho_{21}^{(m)}\) is linearized around its RF-free value \(\rho_{21,0}^{(m)}\) as
\begin{align}
\rho_{21}^{(m)} \approx \rho_{21,0}^{(m)} + \chi_k^{(m)} \Omega_{\mathrm{RF},k}^{(m)}, \label{eq:rho21}
\end{align}
where \(\Omega_{\mathrm{RF},k}^{(m)} = \frac{\mu_{eg}^{T}\epsilon_{m,k}}{\hbar}\sqrt{P_k}\rho_{m,k}|s_k|\) is the RF Rabi frequency induced by user \(k\), with \(P_k\) being the total transmit power of user \(k\), and \(\mu_{eg}\), \(\epsilon_{m,k}\), \(\rho_{m,k}\) denoting the transition dipole moment, polarization, and path loss, respectively. The atomic susceptibility is
\begin{align}
\chi_k^{(m)} &= \left.\frac{\partial \rho_{21}^{(m)}}{\partial \Omega_{\mathrm{RF},k}^{(m)}}\right|_{\Omega_{\mathrm{RF}}=0}
\nonumber\\
& = \frac{i\Omega_p^{(m)}}{2\left[\Gamma_{21}+i\Delta_p^{(m)}+\frac{|\Omega_c^{(m)}|^2}{4\left(\Gamma_{31}+i(\Delta_p^{(m)}+\Delta_c^{(m)})\right)}\right]^2}. \label{eq:chi}
\end{align}

Accordingly, the equivalent channel coefficient from user \(k\) to cell \(m\) is
\begin{align}
h_{m,k} = \eta_{\mathrm{PD}}^{(m)} \chi_k^{(m)} \frac{\mu_{eg}^{T}\epsilon_{m,k}}{\hbar}\sqrt{P_k}\rho_{m,k}, \label{eq:hmk}
\end{align}
with \(\eta_{\mathrm{PD}}^{(m)}\) being the optical readout gain of the \(m\)-th photodetector. Here, the factor \(\sqrt{P_k}\) originates from the fact that the RF Rabi frequency is proportional to the incident field amplitude, which scales with \(\sqrt{P_k}\). In~\eqref{eq:rx-signal}, this factor is absorbed into \(\mathbf{h}_k\), while \(s_k\) carries only the normalized modulation symbol. Finally, stacking \(h_{m,k}\) over all \(M\) cells yields \(\mathbf{h}_k=[h_{1,k},\dots,h_{M,k}]^T\), which shows that \(\mathbf{h}_k\) is jointly determined by \(\{\Omega_p^{(m)},\Omega_c^{(m)},\Delta_p^{(m)},\Delta_c^{(m)},\Gamma_{21},\Gamma_{31},\Gamma_{32},\rho_{21}^{(m)}\}\) and the wireless propagation coefficients \(\{\rho_{m,k},\epsilon_{m,k}\}\). With the atomic origin of \(\mathbf{h}_k\) clarified, we now return to the RAQR observation model.

Unlike traditional RF receivers that directly acquire the complex signal \(\mathbf{y}\), the considered RAQR only measures its amplitude. While coherent detection using lock-in amplifiers \cite{meyer2018digital,simons2019rydberg} is a potential alternative, it is beyond the scope of this work. The amplitude-only observation is therefore expressed as
\begin{align}\label{eq:Observation model}
\mathbf{z} = |\mathbf{y}|
= \left|\sum_{k=1}^{K}\mathbf{h}_k s_k + \mathbf{b} + \mathbf{n}\right|
\in\mathbb{R}^{M\times 1},
\end{align}
where \(\mathbf{z}\) is linked to the Autler--Townes (AT) splitting interval of the underlying Rydberg EIT resonance. This non-linear amplitude relationship complicates signal detection. Nevertheless, \(\mathbf{z}\) can be linearized by establishing phase coherence between the received signal and the strong LO. Provided that the phase alignment condition \(\angle\!\bigl(\sum_{k=1}^{K}\mathbf{h}_k s_k\bigr)\approx \angle\mathbf{b}\) holds and the strong-LO assumption \(|\mathbf{b}|\gg\bigl\|\sum_{k=1}^{K}\mathbf{h}_k s_k\bigr\|_2\) is satisfied, the small-signal approximation yields
\begin{align}\label{eq:simplified form of the signal}
\mathbf{z} \approx |\mathbf{b}| + \sum_{k=1}^{K}\mathbf{h}_k^{\mathrm{opt}}(\psi)\, s_k + \tilde{\mathbf{n}},
\end{align}
where \(\psi=[\psi_1,\ldots,\psi_M]^T\) is the phase alignment vector applied at the \(M\) atomic cells, and \(\mathbf{h}_k^{\mathrm{opt}}(\psi)\) denotes the equivalent real-valued optical channel for user \(k\) under this phase alignment. The equivalent real-valued optical channel is defined as
\begin{align}\label{eq:hopt-psi}
\mathbf{h}_k^{\mathrm{opt}}(\psi)
&=\bigl[\Re\{h_{m,k}e^{j\psi_m}\}\bigr]_{m=1}^{M}
\in \mathbb{R}^{M\times 1}.
\end{align}

In the ideal case where the phase alignment exactly compensates the channel phase of user \(k\), i.e., \(\psi_m=-\angle h_{m,k}\) for all \(m\), we have \(\Re\{h_{m,k}e^{j\psi_m}\}=|h_{m,k}|\), so that \(\mathbf{h}_k^{\mathrm{opt}}(\psi)\) reduces to the elementwise magnitude vector \([\,|h_{1,k}|,\ldots,|h_{M,k}|\,]^T\). However, since a single common phase vector \(\psi\) is shared by all users, this reduction holds simultaneously for all \(k\) only when the channel phases \(\{\angle h_{m,k}\}\) are aligned across users; otherwise, \(\mathbf{h}_k^{\mathrm{opt}}(\psi)\) retains the general form in~\eqref{eq:hopt-psi}, which is adopted in the subsequent optimization. The effective channel gain is therefore \(G_k(\psi)=\|\mathbf{h}_k^{\mathrm{opt}}(\psi)\|_2^2\), consistent with~\eqref{eq:effective_gain}. Precise phase alignment thus transforms the non-linear observation in~\eqref{eq:Observation model} into the linear model in~\eqref{eq:simplified form of the signal}, facilitating efficient user signal recovery and interference cancellation; the detailed derivation and phase alignment algorithm are presented in Section~\ref{sec:Inner_Layer}.

\subsection{Effective Channel Gain and User Ordering}

Based on the linearized observation model in~\eqref{eq:simplified form of the signal}, we define the effective channel gain $G_k(\boldsymbol{\psi})$ as the squared Euclidean norm of the equivalent real-valued channel vector under a specific phase alignment $\boldsymbol{\psi}=[\psi_1,\ldots,\psi_M]^T$:
\begin{align}\label{eq:effective_gain}
G_k(\boldsymbol{\psi}) 
= \bigl\|\mathfrak{R}\{\mathbf{h}_k\circ e^{j\boldsymbol{\psi}}\}\bigr\|^{2}
= \sum_{m=1}^{M}\bigl(\mathfrak{R}\{h_{m,k}\,e^{j\psi_m}\}\bigr)^{2}.
\end{align}

This metric quantifies the aggregate signal combining gain provided by the $M$-element atomic array under a specific phase alignment $\boldsymbol{\psi}$. When phase alignment is achieved, $G_k(\boldsymbol{\psi})$ reduces to $G_k = \sum_{m=1}^{M}|h_{m,k}|^2 = \|\mathbf{h}_k\|_2^2$. Without loss of generality, the users are indexed in non-decreasing order of their effective channel gains:
\begin{align}\label{eq:user_ordering}
G_1 \le G_2 \le G_3.
\end{align}


\subsection{Two Operational Schemes}
\label{sec:two_schemes}

To exploit the potential of Rydberg receivers for uplink RSMA, we consider two complementary transmission schemes. These schemes embody a performance-complexity trade-off: Scheme A (Full RSMA) targets maximum EE through global message splitting and full bandwidth sharing, whereas Scheme B (Paired RSMA) reduces decoding complexity via strong-weak user pairing and orthogonal sub-band allocation, at the expense of partial multiplexing gain.

\paragraph{Scheme~A (Full RSMA)}
All $K=3$ users simultaneously share the entire bandwidth $B$. Each user $k$ splits its message into a common stream $s_k^c$ and a private stream $s_k^p$, which are transmitted with powers $p_k^c$ and $p_k^p$, respectively. The BS performs SIC to decode all $2K$ streams according to a predetermined decoding order. Scheme~A fully exploits the multiplexing gain, which requires a combinatorial decoding-order search.

\paragraph{Scheme~B (Paired RSMA)}
Users are grouped based on a strong-weak pairing strategy \cite{10330667}. Following the user ordering in~\eqref{eq:user_ordering}, the strongest user ($\mathrm{U}_3$) is paired with the weakest ($\mathrm{U}_1$) to form Pair~1, while the median user ($\mathrm{U}_2$) operates as a standalone user. The total bandwidth $B$ is orthogonally divided: Pair~1 is allocated a fraction $\alpha_1 B$, and $\mathrm{U}_2$ is allocated $\alpha_0 B$, subject to $\alpha_0 + \alpha_1 \leq 1$. Within Pair~1, only $\mathrm{U}_1$ splits its message into two sub-streams $x_{11}$ and $x_{12}$ with powers $p_{11}^{\mathrm{B}}$ and $p_{12}^{\mathrm{B}}$, while $\mathrm{U}_3$ transmits a single stream $x_3$ with power $p_3^{\mathrm{B}}$. The SIC decoding order within the pair is fixed as $x_{11} \to x_3 \to x_{12}$ \cite{10330667}. The standalone $\mathrm{U}_2$ transmits over its dedicated orthogonal sub-band without rate splitting, utilizing power $p_2^{\mathrm{B}}$. This orthogonal separation eliminates inter-pair interference, and the deterministic intra-pair decoding order avoids the combinatorial search required in Scheme~A.

\subsection{Scheme~A: Achievable Rates with Full RSMA}
\label{sec:scheme_A_rates}
In Scheme~A, the BS applies maximal ratio combining (MRC) to the phase-aligned real-valued observation in~\eqref{eq:simplified form of the signal}. Specifically, the combining vector for user $k$ is
\begin{align}\label{eq:combining_vector}
\mathbf{w}_k = \frac{\mathbf{h}_k^{\mathrm{opt}}(\boldsymbol{\psi})}{\|\mathbf{h}_k^{\mathrm{opt}}(\boldsymbol{\psi})\|_2} \in \mathbb{R}^{M\times 1},
\end{align}
where $\mathbf{h}_k^{\mathrm{opt}}(\boldsymbol{\psi})$ is defined in~\eqref{eq:hopt-psi}. Applying $\mathbf{w}_k^T$ to $\mathbf{z}$ yields the post-combining observation:
\begin{align}
\mathbf{w}_k^T\mathbf{z} \! \approx \! \mathbf{w}_k^T|\mathbf{b}| \!+\! \underbrace{\mathbf{w}_k^T\mathbf{h}_k^{\mathrm{opt}}(\boldsymbol{\psi})}_{=\,G_k^{1/2}} s_k 
\!+\! \sum_{i\neq k}\mathbf{w}_k^T\mathbf{h}_i^{\mathrm{opt}}(\boldsymbol{\psi})\,s_i \!+\! \mathbf{w}_k^T\tilde{\mathbf{n}}.
\end{align}

Accordingly, the effective channel gain after combining equals \(G_k=\|\mathbf{h}_k^{\mathrm{opt}}(\boldsymbol{\psi})\|_2^2\), which is consistent with \eqref{eq:effective_gain}. The effective real-valued noise after combining is $\tilde{n}_k\triangleq\mathbf{w}_k^T\tilde{\mathbf{n}}$, whose variance is $N_0 \triangleq \mathbb{E}\{\tilde{n}_k^2\} = \frac{\sigma_n^2}{2}$, where the factor $1/2$ arises because the complex baseband noise $\mathbf{n}\sim\mathcal{CN}(\mathbf{0},\sigma_n^2\mathbf{I}_M)$ has independent real and imaginary parts, each with variance $\sigma_n^2/2$, and the real-valued observation model in~\eqref{eq:simplified form of the signal} retains only the real component.

To accommodate heterogeneous channel conditions, the SIC decoding order is specified as follows: common streams are decoded from the strongest user to the weakest (user \(3 \to 2 \to 1\)), since they mutually interfere and are removed sequentially via SIC; private streams are decoded from the weakest user to the strongest (user \(1 \to 2 \to 3\)) after all common streams have been cancelled.

Under this ordering, the post-combining achievable common rates are
\begin{align}
r_3^c &= B\log_2\!\left(\!1 \!+\! \frac{G_3 p_3^c}{G_2 p_2^c \!+\! G_1 p_1^c \!+\! \sum_{i=1}^{K} G_i p_i^p \!+\! N_0}\!\right)\!,\nonumber\\
r_2^c &= B\log_2\!\left(1 + \frac{G_2 p_2^c}{G_1 p_1^c + \sum_{i=1}^{K} G_i p_i^p + N_0}\right),\nonumber\\
r_1^c &= B\log_2\!\left(1 + \frac{G_1 p_1^c}{\sum_{i=1}^{K} G_i p_i^p + N_0}\right)\nonumber.
\end{align}

Similarly, the private rates are
\begin{align}
r_1^p &= B\log_2\!\left(1 + \frac{G_1 p_1^p}{G_2 p_2^p + G_3 p_3^p + N_0}\right),\nonumber\\
r_2^p &= B\log_2\!\left(1 + \frac{G_2 p_2^p}{G_3 p_3^p + N_0}\right),\nonumber\\
r_3^p &= B\log_2\!\left(1 + \frac{G_3 p_3^p}{N_0}\right)\nonumber.
\end{align}

The total achievable rate of user $k$ under Scheme~A is given by $R_k^{\mathrm{A}} = r_k^c + r_k^p$.

\subsection{Scheme~B: Achievable Rates with Paired RSMA}
\label{sec:scheme_B_rates}

In Scheme~B, under the fixed intra-pair SIC decoding order $x_{11} \to x_3 \to x_{12}$, the achievable rates for Pair 1 are:
\begin{align}
    r_{11}^{\mathrm{B}} &= B\alpha_1 \log_2\!\left(1 + \frac{G_1\, p_{11}^{\mathrm{B}}}{G_3\, p_3^{\mathrm{B}} + G_1\, p_{12}^{\mathrm{B}} + N_0}\right), \nonumber\\
    r_3^{\mathrm{B}} &= B\alpha_1 \log_2\!\left(1 + \frac{G_3\, p_3^{\mathrm{B}}}{G_1\, p_{12}^{\mathrm{B}} + N_0}\right), \nonumber\\
    r_{12}^{\mathrm{B}} &= B\alpha_1 \log_2\!\left(1 + \frac{G_1\, p_{12}^{\mathrm{B}}}{N_0}\right), \nonumber
\end{align}
where $G_k$ is the post-combining effective channel gain, and $N_0$ is the effective real-valued noise variance. The same combining and noise definitions apply to Scheme~B. The total rate of $\mathrm{U}_1$ is $R_1^{\mathrm{B}} = r_{11}^{\mathrm{B}} + r_{12}^{\mathrm{B}}$, and the rate of $\mathrm{U}_3$ is $R_3^{\mathrm{B}} = r_3^{\mathrm{B}}$. 

Since $\mathrm{U}_2$ occupies an orthogonal sub-band, it experiences no multi-user interference. Its achievable rate is \(R_2^{\mathrm{B}} = B\alpha_0 \log_2(1 + G_2 p_2^{\mathrm{B}}/N_0)\). Compared with Scheme~A, Scheme~B replaces the combinatorial SIC decoding-order search with a deterministic intra-pair decoding order and continuous bandwidth allocation variables $(\alpha_0, \alpha_1)$.

\subsection{Problem Formulation}
\label{sec:problem_formulation}

Based on the derived achievable rates, we formulate the EE maximization problem for both schemes. The EE is defined as the ratio of the sum rate to the total power consumption:
\begin{align}
\mathrm{EE} = \frac{\sum_{k=1}^{K}R_k}{P_{\mathrm{total}}}, \label{eq:EE_def}
\end{align}
where the total power consumption is $P_{\mathrm{total}} = \sum_{k=1}^{K}(p_k^c + p_k^p) + P_{\mathrm{c}}$,and $P_{\mathrm{c}} = M(P_{\mathrm{vapor}} + P_{\mathrm{PD}}) + P_{\mathrm{DSP}} + P_{\mathrm{LO}}$ represents the hardware power overhead of the RAQR. Although both schemes share the same EE objective, their distinct transmission structures result in different optimization formulations.

\subsubsection{Problem Formulation for Scheme~A}
\label{sec:prob_A}

Using the EE definition in \eqref{eq:EE_def}, the joint power allocation and phase alignment optimization problem for Scheme~A is formulated as:
\begin{subequations}\label{prob:scheme_A}
\begin{align}
(\mathcal{P}_{\mathrm{A}}):\
\max_{\{p_k^c,p_k^p\geq 0\},\boldsymbol{\psi}}\quad & \mathrm{EE}^{\mathrm{A}} = \frac{\sum_{k=1}^{K}(r_k^c + r_k^p)}{\sum_{k=1}^{K}(p_k^c + p_k^p) + P_{\mathrm{c}}} \label{eq:obj_EE}\\
\text{s.t.}\quad
& p_k^c + p_k^p \le P_k^{\max},\ \forall k, \label{eq:C1_power_budget}\\
& r_k^c(\boldsymbol{\psi}) + r_k^p(\boldsymbol{\psi}) \ge R_k^{\min},\ \forall k, \label{eq:C2_QoS}\\
& \bigl\|\mathfrak{I}\{\mathbf{H}^{\mathrm{eq}}(\boldsymbol{\psi})\}\bigr\|_F^2 \le \epsilon, \label{eq:C3_phase}
\end{align}
\end{subequations}
where $\mathbf{H}^{\mathrm{eq}}(\boldsymbol{\psi}) = [\mathbf{h}_1\circ e^{j\boldsymbol{\psi}},\,\mathbf{h}_2\circ e^{j\boldsymbol{\psi}},\,\mathbf{h}_3\circ e^{j\boldsymbol{\psi}}]$ is the composite effective channel matrix, and $\circ$ denotes the Hadamard product. Constraint~\eqref{eq:C1_power_budget} ensures that each user's transmit power is within the hardware budget $P_k^{\max}$. Constraint~\eqref{eq:C2_QoS} guarantees that the achievable rate meets the individual minimum-rate requirement $R_k^{\min}$. Constraint~\eqref{eq:C3_phase} maintains phase alignment by restricting the energy of the residual imaginary component below a small tolerance $\epsilon > 0$, thereby ensuring the validity of the linear model in~\eqref{eq:simplified form of the signal}.

\subsubsection{Problem Formulation for Scheme~B}
\label{sec:prob_B}

For Scheme~B, the optimization variables include the transmit powers $\{p_{11}^{\mathrm{B}}, p_{12}^{\mathrm{B}}, p_3^{\mathrm{B}}, p_2^{\mathrm{B}}\}$, the bandwidth allocation factors $\{\alpha_0, \alpha_1\}$, and the phase alignment vector $\boldsymbol{\psi}$. The problem is formulated as:
\begin{subequations}\label{prob:scheme_B}
\begin{align}
(\mathcal{P}_{\mathrm{B}}):\quad
\max_{\substack{p_{11}^{\mathrm{B}},\, p_{12}^{\mathrm{B}},\, p_3^{\mathrm{B}},\, p_2^{\mathrm{B}} \geq 0,\\[2pt] \alpha_0,\,\alpha_1 > 0,\;\boldsymbol{\psi}}} \quad 
&\!\!\!\! \mathrm{EE}^{\mathrm{B}} \!=\! \frac{R_1^{\mathrm{B}} \!+\! R_2^{\mathrm{B}} \!+\! R_3^{\mathrm{B}}}
{p_{11}^{\mathrm{B}} \!+\! p_{12}^{\mathrm{B}} \!+\! p_2^{\mathrm{B}} \!+\! p_3^{\mathrm{B}} \!+\! P_{\mathrm{c}}} \label{eq:obj_B} \\
\text{s.t.} \quad 
& p_{11}^{\mathrm{B}} + p_{12}^{\mathrm{B}} \leq P_1^{\max}, \label{eq:power1_B} \\
& p_2^{\mathrm{B}} \leq P_2^{\max}, \label{eq:power2_B} \\
& p_3^{\mathrm{B}} \leq P_3^{\max}, \label{eq:power3_B} \\
& R_k^{\mathrm{B}} \geq R_k^{\min}, \quad k \!=\! 1,2,3, \label{eq:qos_B} \\
& \alpha_0 + \alpha_1 \leq 1, \label{eq:bw_B} \\
& \bigl\|\mathfrak{I}\{\mathbf{H}^{\mathrm{eq}}(\boldsymbol{\psi})\}\bigr\|_F^2 \leq \epsilon. \label{eq:phase_B}
\end{align}
\end{subequations}

Here, constraints~\eqref{eq:power1_B}--\eqref{eq:power3_B} impose individual power budgets, \eqref{eq:qos_B} guarantees the rate requirements, \eqref{eq:bw_B} enforces the orthogonal bandwidth partitioning, and \eqref{eq:phase_B} maintains the phase alignment validity identical to constraint~\eqref{eq:C3_phase}.

\subsubsection{Non-Convexity Analysis}

Both $\mathcal{P}_{\mathrm{A}}$ and $\mathcal{P}_{\mathrm{B}}$ are non-convex due to four factors: (i) the non-convex phase alignment constraint $\|\mathcal{I}\{\mathbf{H}_{eq}(\boldsymbol{\psi})\}\|_F^2 \leq \varepsilon$, which involves trigonometric functions of $\boldsymbol{\psi}$ and couples with the power allocation variables; (ii) the fractional structure of the EE objective; (iii) the logarithmic rate expressions involving multi-user interference; and (iv) the coupling between the phase vector $\boldsymbol{\psi}$ and the power variables. Compared to $\mathcal{P}_{\mathrm{A}}$, problem $\mathcal{P}_{\mathrm{B}}$ reduces the number of power variables and eliminates the combinatorial SIC decoding-order search by introducing bandwidth allocation variables. While $\mathcal{P}_{\mathrm{A}}$ yields superior EE by fully exploiting spatial and multiplexing gains, $\mathcal{P}_{\mathrm{B}}$ enables low-complexity resource allocation.

\section{Proposed Unified Three-Layer Decomposition Framework}
\label{sec:framework}
To address the non-convexity of Problems~\eqref{prob:scheme_A} and~\eqref{prob:scheme_B}, we propose a unified hierarchical three-layer optimization framework tailored to the RAQR constraints. The framework systematically decomposes both schemes as follows:

\begin{enumerate}
    \item \textbf{Inner Layer (Phase Alignment -- Common to Both Schemes):} Given the current power allocation, this layer estimates the phase compensation parameters $\{\theta_m\}$ to satisfy the phase alignment constraints~\eqref{eq:C3_phase} or~\eqref{eq:phase_B} using an extended EM-GS algorithm. This process ensures a valid real-valued channel linearization for both schemes.
    
    \item \textbf{Middle Layer (Resource Allocation -- Scheme-Specific):}
    \begin{itemize}
        \item \textit{Scheme A (Full RSMA):} Based on the aligned phases, the power variables $\{p_k^c, p_k^p\}$ are optimized via closed-form solutions in symmetric scenarios or block coordinate descent power allocation (BCD-PA) in asymmetric scenarios, as detailed in Section~\ref{sec:middle_A}.
        \item \textit{Scheme B (Paired RSMA):} Given the aligned phases, the power $\{p_{11}^m, p_{12}^m, p_2^m\}$ and bandwidth allocation $\{\alpha_m\}$ are jointly optimized via closed-form expressions and bisection search, as detailed in Section~\ref{sec:middle_B}.
    \end{itemize}
    
    \item \textbf{Outer Layer (EE Maximization -- Common to Both Schemes):} The Dinkelbach parameter $\eta$ is iteratively updated based on the rate and power allocation obtained from the middle layer. The Dinkelbach iteration structure is identical for both schemes, differing only in the specific rate and power expressions. Upon convergence, minimum mean square error (MMSE) detection is performed for final signal recovery.
\end{enumerate}

This unified framework exploits the underlying structural commonality of the system: both schemes require phase alignment (due to the RAQR hardware constraints) and EE maximization, differing only in the multi-user spectrum access mechanism (full sharing versus orthogonal pairing).

\subsection{Inner Layer: Phase Alignment via Extended EM-GS Algorithm}
\label{sec:Inner_Layer}
The amplitude-only measurement of the RAQR introduces a biased phase retrieval (PR) problem, as the phase information of the received signal is lost in the presence of a known local-oscillator (LO) field $\mathbf{b}$. This mathematical structure is equivalent to the biased PR model established in \cite{cui2025towards} for atomic MIMO receivers. Consequently, we adopt the expectation-maximization Gerchberg-Saxton (EM-GS) framework and extend it with an explicit gradient-descent step to optimize the phase compensation vector $\boldsymbol{\psi}$ under the linearization constraint of the RAQR.

For each user $k$, the total transmit power is $p_k = p_k^c + p_k^p$. The transmitted composite symbol is given by \eqref{eq:tx_signal}, where $x_k^c$ and $x_k^p$ are independent, unit-power symbols. For phase alignment and channel estimation, the normalized composite symbol is defined as \(x_k = s_k/\sqrt{p_k} = (\sqrt{p_k^c}\,x_k^c + \sqrt{p_k^p}\,x_k^p)/\sqrt{p_k}\), yielding $s_k = \sqrt{p_k}\, x_k$ with $\mathbb{E}[|x_k|^2]=1$. This representation is mathematically consistent with the RSMA signal model, as the RAQR measures the aggregate field $\sum_k \mathbf{h}_k s_k$ without separating the individual data streams at the physical layer.

At the $m$-th vapor cell, the amplitude observation follows directly from \eqref{eq:Observation model} with the total symbol vector $\mathbf{s}=[\sqrt{p_1}x_1,\dots,\sqrt{p_K}x_K]^T$, i.e., $\mathbf{z}=|\mathbf{A}^H\mathbf{s}+\mathbf{b}+\mathbf{n}|$, where $\mathbf{A}$ is the equivalent channel matrix and $\mathbf{b}$ is the known LO field. This form aligns exactly with the biased PR model in \cite{cui2025towards}.

It is important to clarify how the circular dependence between the EM-GS phase retrieval and symbol detection is broken in practice. The proposed inner-layer algorithm operates in two stages: a pilot-based initialization stage and a data-aided iterative stage.

During the pilot phase, each user $k$ transmits a known pilot sequence $\{x_k^{\mathrm{p}}\}$ with a predefined power. Stacking the pilot symbols from all users into $\mathbf{s}^{\mathrm{p}} = [\sqrt{p_1}x_1^{\mathrm{p}},\ldots,\sqrt{p_K}x_K^{\mathrm{p}}]^T$, the RAQR collects $N_{\mathrm{p}}$ amplitude observations $\{\mathbf{z}_n^{\mathrm{p}}\}_{n=1}^{N_{\mathrm{p}}}$ according to~\eqref{eq:Observation model}. Since the pilots are known, the initial total-symbol estimate \(\hat{\mathbf{s}}^{(0)}\) can be obtained by the pilot-based least-squares solution
\begin{align}
\hat{\mathbf{s}}^{(0)} = \bigl(\mathbf{A}^H\mathbf{A}\bigr)^{-1}\mathbf{A}^H\bigl(\mathbf{z}^{\mathrm{p}}\circ e^{j\angle\mathbf{b}} - \mathbf{b}\bigr).\label{eq:pilot_init}
\end{align}

Alternatively, \(\hat{\mathbf{s}}^{(0)}\) can be obtained by the spectral initialization method in~\cite{cui2025towards}, which uses the principal eigenvector of the augmented covariance matrix \(\mathbf{M}=\sum_{n=1}^{N_{\mathrm{p}}}z_n^{\mathrm{p}}\bar{\mathbf{a}}_n\bar{\mathbf{a}}_n^H\) to obtain a coarse estimate without requiring any prior symbol knowledge. This removes the need for data symbols in the initialization of the EM-GS iteration.

Once $\hat{\mathbf{s}}^{(0)}$ is available, the EM-GS iteration alternates between estimating the latent phase $\theta_m$ and refining the total-symbol estimate $\hat{\mathbf{s}}^{(i)}$. At each iteration, $\hat{\mathbf{s}}^{(i)}$ is not a hard decision on the transmitted symbols; it is a soft, linear least-squares estimate obtained from the previous M-step. Therefore, the iteration does not require the final symbol decisions and can be carried out before demodulation.

After the EM-GS iteration converges, the phase alignment vector $\boldsymbol{\psi}^*$ is fixed and the equivalent real-valued channel $\bar{h}_{m,k}$ is obtained. The subsequent resource allocation and MMSE detection then recover the data symbols. In other words, the phase retrieval stage uses only pilots (for initialization) and soft symbol estimates (for iteration), while the data symbols are detected only after the phase alignment is completed.

Given the current total-symbol estimate $\hat{\mathbf{s}}^{(i)}$, the conditional distribution of the observed amplitude $z_m$ is Rician. The posterior distribution of the latent phase $\theta_m \triangleq \angle\bigl(\mathbf{a}_m^H \mathbf{s} + b_m\bigr)$ follows a von Mises distribution:
\begin{align}
p\bigl(\theta_m \mid z_m, \hat{\mathbf{s}}^{(i)}\bigr) = \frac{\exp\Bigl(\kappa_m^{(i)}\cos\bigl(\theta_m - \mu_m^{(i)}\bigr)\Bigr)}{2\pi I_0\bigl(\kappa_m^{(i)}\bigr)},
\end{align}
where $\mu_m^{(i)} = \angle\bigl(\mathbf{a}_m^H \hat{\mathbf{s}}^{(i)} + b_m\bigr)$ represents the posterior phase mean, and $\kappa_m^{(i)} = \dfrac{2z_m \bigl|\mathbf{a}_m^H \hat{\mathbf{s}}^{(i)} + b_m\bigr|}{\sigma^2}$ is the concentration parameter.

The E-step produces \(\tilde{\mathbf{z}}^{(i)} = \mathbf{z} \circ e^{j\boldsymbol{\mu}^{(i)}} \circ R(\boldsymbol{\kappa}^{(i)})\), where $R(x)=I_1(x)/I_0(x)$ is the ratio of the first-kind modified Bessel functions, and $\circ$ denotes the Hadamard product. As established in \cite{cui2025towards}, $R(x)$ functions as an SNR-dependent high-pass filter, suppressing unreliable phase estimates under low-SNR conditions while converging to $1$ at high SNR.

Using $\tilde{\mathbf{z}}^{(i)}$ as the effective linear observation, the M-step updates \(\hat{\mathbf{s}}^{(i+1)} = (\mathbf{A}^H\mathbf{A})^{-1}\mathbf{A}^H(\tilde{\mathbf{z}}^{(i)} - \mathbf{b})\). This closed-form update maximizes the expected log-likelihood and is identical to the standard EM-GS algorithm.

To minimize the residual imaginary energy of the equivalent channel for RAQR linearization, we augment the EM-GS iteration with a gradient descent step on the phase compensation vector $\boldsymbol{\psi}$. We define the cost function penalizing the residual imaginary components as
\begin{align}\label{eq:cost_J_2}
J(\boldsymbol{\psi}) = \sum_{m=1}^{M} \sum_{k=1}^{K} \Bigl| \Im\bigl\{ h_{m,k} e^{j\psi_k} \bigr\} \Bigr|^2,
\end{align}
where $\psi_k$ denotes the phase compensation applied to user $k$'s channel across all vapor cells. The gradient with respect to $\psi_k$ is given by
\begin{align}\label{eq:grad_J}
\frac{\partial J}{\partial \psi_k} = 2 \sum_{m=1}^{M} \Im\bigl\{ h_{m,k} e^{j\psi_k} \bigr\} \cdot \Re\bigl\{ h_{m,k} e^{j\psi_k} \bigr\}.
\end{align}

Following each EM-GS iteration, the phase vector is updated via \(\boldsymbol{\psi}^{(i+1)} = \boldsymbol{\psi}^{(i)} - \mu \nabla J(\boldsymbol{\psi}^{(i)})\), where $\mu>0$ is the step size. This extension ensures adherence to the RAQR linearization requirements while maintaining the theoretical advantages of the EM framework. 


\subsection{Middle Layer for Scheme A: Power Allocation}\label{sec:middle_A}

Given the phase-aligned real-valued channel matrix $\tilde{\mathbf{H}}$ with entries $\bar{h}_{m,k} = \Re\{h_{m,k}e^{j\psi_k^*}\}$ from the inner layer, the effective channel gain for user $k$ is defined as $G_k = \sum_{m=1}^{M} \bar{h}_{m,k}^2$. Without loss of generality, we order the users such that $G_1 \leq G_2 \leq \cdots \leq G_K$. The middle layer optimizes the power allocation based on these gains, distinguishing between symmetric and asymmetric rate-requirement scenarios.

\subsubsection{Symmetric Rate-Requirement Scenario: Closed-Form 
Solution}\label{sec:symmetric}
For homogeneous deployments where users share similar rate requirements, the Symmetric Rate-Requirement condition is quantified as:
\begin{align}\label{eq:sym_condition}
    \max_{k} \frac{|R_k^{\min} - \bar{R}|}{\bar{R}} < \delta,
\end{align}
where $\bar{R} = \frac{1}{K}\sum_{k=1}^K R_k^{\min}$ and $\delta$ is a predefined threshold (e.g., $0.1$). When~\eqref{eq:sym_condition} is met, a common target rate $R$ is enforced for all users with rate-splitting factors satisfying $\alpha + \beta = 1$. Note that symmetric rate requirement does not imply symmetric channel gains; the derived closed-form solution is valid for arbitrarily ordered channel gains $G_1 \leq G_2 \leq G_3$.  It is worth clarifying why the rate-splitting factors $\alpha$ and $\beta$ are treated as fixed parameters rather than optimization variables in Problem~\eqref{prob:scheme_A}. 

In the considered symmetric-rate scenario, all users share the same target rate $R$, and $\alpha+\beta=1$ only determines how $R$ is split between the common and private streams. Once $R$ is fixed, any feasible pair $(\alpha,\beta)$ yields the same total rate $R$; the split affects neither the sum rate nor the total transmit power required to achieve $R$, and hence does not change the achievable EE. Including $\alpha$ and $\beta$ as additional optimization variables would therefore introduce a redundancy without improving the objective. Moreover, the rate-splitting factors are typically chosen offline according to the RSMA protocol design and the target rate ratio between common and private streams \cite{10330667}; fixing them reduces the dimensionality of the optimization and enables the closed-form KKT solution in~\eqref{eq:closed_form_private}--\eqref{eq:closed_form_common}. 

Assuming a three-user scenario ($K=3$) with standard SIC ordering (weakest user decoded first for private streams), applying the KKT conditions under the equal-rate constraint $r_k^p = \beta R$ yields the following closed-form optimal private power allocation:
\begin{subequations}\label{eq:closed_form_private}
\begin{align}
    p_3^{p*} &= \frac{N_0}{G_3}\bigl(2^{\beta R / B} - 1\bigr), \label{eq:p3_private}\\
    p_2^{p*} &= \frac{N_0}{G_2}\bigl(2^{2\beta R / B} - 2^{\beta R / B}\bigr), \label{eq:p2_private}\\
    p_1^{p*} &= \frac{N_0}{G_1}\bigl(2^{3\beta R / B} - 2^{2\beta R / B}\bigr). \label{eq:p1_private}
\end{align}
\end{subequations}

For the common streams decoded in reverse order ($3 \to 2 \to 1$), we define $\gamma_{\mathrm{th}} = 2^{\alpha R / B} - 1$ and the interference floor as $I = \sum_{i=1}^{3} G_i p_i^{p*} + N_0$. The optimal common power allocation is derived as:
\begin{subequations}\label{eq:closed_form_common}
\begin{align}
    p_1^{c*} &= \frac{I}{G_1}\,\gamma_{\mathrm{th}}, \label{eq:p1_common}\\
    p_2^{c*} &= \frac{I}{G_2}\,\gamma_{\mathrm{th}}\bigl(\gamma_{\mathrm{th}} + 1\bigr), \label{eq:p2_common}\\
    p_3^{c*} &= \frac{I}{G_3}\,\gamma_{\mathrm{th}}\bigl(\gamma_{\mathrm{th}} + 1\bigr)^2. \label{eq:p3_common}
\end{align}
\end{subequations}

The structural recursion scales identically for arbitrary $K$. For a fixed rate $R$, the EE is defined as
\begin{align}\label{eq:EE_R_sym}
    \mathrm{EE}(R) = \frac{3R}{\sum_{k=1}^{3} \bigl(p_k^{c*}(R) + p_k^{p*}(R)\bigr) + P_{\mathrm{c}}}.
\end{align}

The unimodality of $\mathrm{EE}(R)$ over the interval $[R_{\min}, R_{\max}]$ is established as follows. The total transmit power $P_{\text{tx}}(R) = \sum_{k}\bigl(p_k^{c*}(R)+p_k^{p*}(R)\bigr)$ is a positive linear combination of exponential functions $2^{\beta R/B}$ and $2^{\alpha R/B}$. Consequently, its logarithmic derivative $h(R) = \frac{P_{\text{tx}}'(R)}{P_{\text{tx}}(R)+P_c}$ is strictly monotonically increasing. Since $\frac{d}{dR}\ln \mathrm{EE}(R) = \frac{1}{R} - h(R)$, the derivative $\frac{d}{dR}\ln \mathrm{EE}(R)$ is positive for small $R$ and negative for large $R$, guaranteeing a unique maximizer $R^*$. 

\subsubsection{Asymmetric-Rate Scenario: BCD-PA Solution}

When the symmetric condition~\eqref{eq:sym_condition} is violated, the non-convex EE maximization problem is solved using BCD-PA. The fractional program is \(\max_{\mathbf{p}\in\mathcal{F}} R(\mathbf{p})/P(\mathbf{p})\), where \(R(\mathbf{p})=\sum_k(r_k^c+r_k^p)\) and \(P(\mathbf{p})=\sum_k(p_k^c+p_k^p)+P_c\).

Applying Dinkelbach's transform, the optimal EE $\eta^*$ satisfies $\max_{\mathbf{p} \in \mathcal{F}} \bigl[ R(\mathbf{p}) - \eta^* P(\mathbf{p}) \bigr] = 0$. For a given parameter $\eta$, the subproblem objective is
\begin{align}
F(\eta) = \max_{\mathbf{p} \in \mathcal{F}} \left[ \sum_{k=1}^{K} (r_k^c + r_k^p) - \eta \left( \sum_{k=1}^{K} (p_k^c + p_k^p) + P_c \right) \right].
\label{eq:subproblem_objective}
\end{align}

Due to the coupled power variables inside the logarithmic rate expressions,~\eqref{eq:subproblem_objective} is non-convex. We apply block coordinate descent, updating one power variable at a time while fixing the others, thereby decomposing the problem into a sequence of convex single-variable subproblems.

For private streams decoded in weakest-first order ($1 \to 2 \to 3$), the SINR for user $k$ is $\mathrm{SINR}_k^p = \frac{G_k p_k^p}{I_k^p}$, where the effective interference is \(I_3^p=N_0\), \(I_2^p=G_3p_3^p+N_0\), and \(I_1^p=G_2p_2^p+G_3p_3^p+N_0\).

Setting the partial derivative of the objective with respect to $p_k^p$ to zero yields the unconstrained optimum $p_k^{p,\text{unc}} = \frac{B}{\eta \ln 2} - \frac{I_k^p}{G_k}$. Projecting this onto the feasible region yields the update rule:
\begin{align}
p_k^p \leftarrow \min\!\left( \max\!\left(0,\; \frac{B}{\eta \ln 2} - \frac{I_k^p}{G_k}\right),\; P_k^{\max} - p_k^c \right). \label{eq:update_pp}
\end{align}

Similarly, for common streams decoded in strongest-first order ($3 \to 2 \to 1$), the SINR is $\mathrm{SINR}_k^c = \frac{G_k p_k^c}{I_k^c}$, where the interference $I_k^c$ is $I_3^c = G_2 p_2^c + G_1 p_1^c + \sum_{i=1}^{K} G_i p_i^p + N_0$, $I_2^c = G_1 p_1^c + \sum_{i=1}^{K} G_i p_i^p + N_0$ and $I_1^c = \sum_{i=1}^{K} G_i p_i^p + N_0$.

The optimal common power update is similarly obtained via projection:
\begin{align}
p_k^c \leftarrow \min\!\left( \max\!\left(0,\; \frac{B}{\eta \ln 2} - \frac{I_k^c}{G_k}\right),\; P_k^{\max} - p_k^p \right). \label{eq:update_pc}
\end{align}

\subsection{Middle Layer for Scheme B: Joint Power and Bandwidth Allocation}\label{sec:middle_B}

Given the phase-aligned channel gains $\{\bar{G}_1^m, \bar{G}_2^m\}$ and the Dinkelbach parameter $\eta$, the middle layer for Scheme B maximizes the decoupled objective:
\begin{align}
F_B(\eta) & = \max_{\{\alpha_m, \mathbf{p}^m\} \in \mathcal{F}_B} \Biggl[\sum_{m=1}^{M}(r_1^m + r_2^m) \nonumber \\
& \qquad - \eta\Biggl(\sum_{m=1}^{M}(p_{11}^m + p_{12}^m + p_2^m) + P_c\Biggr)\Biggr]. \label{eq:dinkel_B}
\end{align}

For a specific bandwidth fraction $\alpha_m$, the achievable rate region $(r_1^m,r_2^m)$ of pair $m$ is bounded by three line segments:
\begin{subequations}\label{eq:rate_region_B}
\begin{align}
\text{AB: } & p_{12}^m = 0, \quad p_2^m = P_{\max,2}^m, \quad 0 \leq p_{11}^m \leq P_{\max,1}^m, \label{eq:seg_AB}\\
\text{BC: } & p_{11}^m + p_{12}^m = P_{\max,1}^m, \quad p_2^m = P_{\max,2}^m, \label{eq:seg_BC}\\
\text{CD: } & p_{11}^m = 0, \quad p_{12}^m = P_{\max,1}^m, \quad 0 \leq p_2^m \leq P_{\max,2}^m. \label{eq:seg_CD}
\end{align}
\end{subequations}

The corresponding maximum rates and sum-rate are
\begin{align}
r_2^{m,\max} &= B\alpha_m \log_2\!\left(1 + \frac{\bar{G}_2^m P_{\max,2}^m}{N_0}\right),\nonumber\\
r_1^m + r_2^m &= B\alpha_m \log_2\!\left(1 + \frac{\bar{G}_1^m P_{\max,1}^m + \bar{G}_2^m P_{\max,2}^m}{N_0}\right),\nonumber\\
r_1^{m,\max} &= B\alpha_m \log_2\!\left(1 + \frac{\bar{G}_1^m P_{\max,1}^m}{N_0}\right)\nonumber.
\end{align}

Under Dinkelbach's transform, optimal rate allocation aligns with the rate target ratio when constraints are active, i.e., \(r_1^m/r_2^m = R_{\min,1}^m/R_{\min,2}^m \triangleq \rho_m\).
The closed-form power allocation for the corresponding operating segments is given by:

Case 1 (on Segment AB):
\begin{align}
p_{11}^m = \frac{\left(2^{\frac{R_{\min,1}^m}{B\alpha_m^{AB}}} - 1\right)\left(\bar{G}_2^m P_{\max,2}^m + N_0\right)}{\bar{G}_1^m},\nonumber\\
\quad p_{12}^m = 0, \quad p_2^m = P_{\max,2}^m.
\label{eq:case1_power}
\end{align}

Case 2 (on Segment BC):
\begin{align}
p_{12}^m &= \frac{\bar{G}_2^m P_{\max,2}^m}{\bar{G}_1^m\left(2^{\frac{R_{\min,2}^m}{B\alpha_m^{BC}}} - 1\right)} - \frac{N_0}{\bar{G}_1^m},\nonumber\\
p_{11}^m &= P_{\max,1}^m - p_{12}^m, \quad p_2^m = P_{\max,2}^m. \label{eq:case2_p11}
\end{align}

Case 3 (on Segment CD):
\begin{align}
p_{12}^m = P_{\max,1}^m, \quad p_{11}^m = 0, \nonumber\\
\quad p_2^m = \frac{\left(2^{\frac{R_{\min,2}^m}{B\alpha_m^{CD}}} - 1\right)\left(\bar{G}_1^m P_{\max,1}^m + N_0\right)}{\bar{G}_2^m}.
\label{eq:case3_power}
\end{align}

The EE-optimal bandwidth fraction $\alpha_m$ can be obtained via a bisection search over $(0,1)$ to solve the KKT condition. For instance, in Case 1, the KKT condition is \(\partial[(r_1^m + r_2^m) - \eta(p_{11}^m + p_{12}^m + p_2^m)]/\partial\alpha_m = 0\).Alternatively, when rate constraints are exactly met, the required bandwidth admits explicit solutions via the Lambert-W function $W(\cdot)$:

Case 1:
\begin{align}\label{eq:alpha_AB}
\alpha_m^{AB} = \frac{R_{\min,2}^m \ln 2}{B\left[W\!\left(\frac{\bar{G}_2^m P_{\max,2}^m + N_0}{N_0} \cdot \ln 2 \cdot e^{\frac{R_{\min,2}^m \ln 2}{B} - 1}\right) + 1\right]}.
\end{align}

Case 2 (with $\Gamma_m^{BC} = \frac{\bar{G}_1^m P_{\max,1}^m + \bar{G}_2^m P_{\max,2}^m}{N_0}$ and $\bar{R}_m = R_{\min,1}^m + R_{\min,2}^m$):
\begin{align}\label{eq:alpha_BC}
\alpha_m^{BC} = \frac{\bar{R}_m \ln 2}{B\left[W\!\left(\Gamma_m^{BC} \ln 2 \cdot e^{\frac{\bar{R}_m \ln 2}{B} - 1}\right) + 1\right]}.
\end{align}

Case 3:
\begin{align}\label{eq:alpha_CD}
\alpha_m^{CD} = \frac{R_{\min,1}^m \ln 2}{B\left[W\!\left(\frac{\bar{G}_1^m P_{\max,1}^m + N_0}{N_0} \cdot \ln 2 \cdot e^{\frac{R_{\min,1}^m \ln 2}{B} - 1}\right) + 1\right]}.
\end{align}

The optimal case index is selected by maximizing the per-pair EE over the feasible set $\mathcal{I}_{\text{feas}} \subseteq \{1, 2, 3\}$:
\begin{align}\label{eq:case_select_EE}
\text{Case}^* = \arg\max_{i \in \mathcal{I}_{\text{feas}}} \; 
\frac{R_m^{(i)}}{P_m^{(i)} + \alpha_m^{(i)} P_{c,m}}.
\end{align}

Overall bandwidth feasibility $\sum_{m=1}^{M} \alpha_m^* \leq 1$ is maintained by adjusting the rate targets dynamically if required.
\vspace{0.5em}
\hrule
\vspace{0.5em}
\begin{algorithm}\label{alg:middle_layer}
    Middle-Layer Resource Allocation for Scheme A and Scheme B
    \vspace{0.5em}
    \hrule
    \vspace{0.5em}
    \begin{enumerate}
        \item \textbf{Input}: Phase-aligned real-valued channel $\tilde{\mathbf{H}}$, channel gains $\{G_k\}$, Dinkelbach parameter $\eta$, scheme selection flag $\mathcal{S} \in \{A, B\}$, rate requirements $\{R_k^{\min}\}$, power bounds $\{P_k^{\max}\}$, tolerance $\varepsilon_{\mathrm{mid}}$.

        \item \textbf{Output}: Resource allocation: Scheme A: $\{p_k^{c*}, p_k^{p*}\}$; Scheme B: $\{p_{11}^{m*}, p_{12}^{m*}, p_2^{m*}, \alpha_m^*\}$.

        \item \textbf{If} $\mathcal{S} = A$:
        \begin{enumerate}
            \item \textbf{Scheme A: Full RSMA Power Allocation}
            \begin{enumerate}
                \item Compute $G_k = \sum_{m=1}^{M} \bar{h}_{m,k}^2$, and sort $G_1 \le G_2 \le G_3$.
                \item Evaluate rate-requirement symmetry metric:
                $\delta_{\mathrm{sym}} = \max_k \frac{|R_k^{\min} - \bar{R}|}{\bar{R}}.$
                \item \textbf{If} $\delta_{\mathrm{sym}} < 0.1$:
                \begin{enumerate}
                    \item Determine $R_{\min} = \max_k R_k^{\min}$.
                    \item Find $R_{\max}$ via feasibility bisection ensuring $p_k^{c*} + p_k^{p*} \le P_k^{\max}$, $\forall k$.
                    \item Initialize $a \gets R_{\min}$, $b \gets R_{\max}$, $\varphi = (\sqrt{5}-1)/2$.
                    \item Compute $c = b - \varphi(b-a)$, $d = a + \varphi(b-a)$.
                    \item \textbf{While} $(b-a) > \epsilon_R$:
                    \begin{itemize}
                        \item Compute $\mathrm{EE}(c)$ and $\mathrm{EE}(d)$ using the closed-form private powers in \eqref{eq:closed_form_private}, common powers in \eqref{eq:closed_form_common}, and $\mathrm{EE}(R)$ in \eqref{eq:EE_R_sym}.
                        \item \textbf{If} $\mathrm{EE}(c) > \mathrm{EE}(d)$: $b \gets d$, $d \gets c$, $c \gets b - \varphi(b-a)$.
                        \item \textbf{Else}: $a \gets c$, $c \gets d$, $d \gets a + \varphi(b-a)$.
                    \end{itemize}
                    \item Set $R^* \gets (a+b)/2$.
                    \item Compute final powers $\{p_k^{c*}\}$, $\{p_k^{p*}\}$ at $R^*$.
                \end{enumerate}
                \item \textbf{Else}:
                \begin{enumerate}
                    \item Initialize $p_k^c = p_k^p = 0.25 P_k^{\max}$ for all $k$.
                    \item \textbf{Repeat}:
                    \begin{itemize}
                        \item \textbf{Stage 1: Private Power Update} (order $k = K, \ldots, 1$).
                        \begin{itemize}
                            \item Compute interference: $I_k^p = \sum_{i>k} G_i p_i^p + N_0$.
                            \item Update $p_k^p$ via \eqref{eq:update_pp}.
                            \item If $r_k^p < R_k^{\min} - r_k^c$, increase $p_k^p$ to meet the rate constraint.
                        \end{itemize}
                        \item \textbf{Stage 2: Common Power Update} (order $k = K, \ldots, 1$).
                        \begin{itemize}
                            \item Compute interference: $I_k^c = \sum_{j<k} G_j p_j^c + \sum_{i=1}^{K} G_i p_i^p + N_0$.
                            \item Update $p_k^c$ via \eqref{eq:update_pc}.
                            \item If $r_k^c < R_k^{\min} - r_k^p$, increase $p_k^c$ to meet the rate constraint.
                        \end{itemize}
                    \end{itemize}
                    \item \textbf{Until} $\|\mathbf{p}^{(t+1)} - \mathbf{p}^{(t)}\|_2 < \varepsilon_{\mathrm{mid}}$.
                    \item Output $\{p_k^{c*}, p_k^{p*}\}$.
                \end{enumerate}
            \end{enumerate}
        \end{enumerate}

        \item \textbf{Else If} $\mathcal{S} = B$:
        \begin{enumerate}
            \item \textbf{Scheme B: Paired RSMA Joint Power and Bandwidth Allocation}
            \begin{enumerate}
                \item \textbf{For} each pair $m = 1, \ldots, M$:
                \begin{enumerate}
                    \item Construct the rate region with segments AB, BC, CD.
                    \item For each feasible case $i \in \{1, 2, 3\}$, compute closed-form powers via \eqref{eq:case1_power}--\eqref{eq:case3_power} and bandwidth via \eqref{eq:alpha_AB}--\eqref{eq:alpha_CD}.
                    \item Select the best case by maximizing per-pair EE:
                    $ \text{Case}^* = \arg\max_{i \in \mathcal{I}_{\text{feas}}} \; \frac{R_m^{(i)}}{P_m^{(i)} + \alpha_m^{(i)} P_{c,m}}.$
                    \item If needed, refine $\alpha_m$ by bisection search on the KKT condition.
                \end{enumerate}
                \item Enforce overall bandwidth feasibility $\sum_{m=1}^{M} \alpha_m^* \le 1$ by adjusting rate targets if necessary.
            \end{enumerate}
            \item Output $\{p_{11}^{m*}, p_{12}^{m*}, p_2^{m*}, \alpha_m^*\}$.
        \end{enumerate}
    \end{enumerate}
    \vspace{0.5em}
    \hrule
    \vspace{0.5em}
\end{algorithm}

\subsection{Outer Layer: Unified Dinkelbach Iteration}

The outer layer performs the generic Dinkelbach parameter update for both schemes. At iteration $t$, the update rule is: \(\eta^{(t+1)} = R_{\text{total}}^{(t)}/P_{\text{total}}^{(t)}\), where the scheme-specific rate and power aggregates are defined as: for Scheme A, \(R_{\text{total}}^{(t)}=\sum_k(r_k^c+r_k^p)\) and \(P_{\text{total}}^{(t)}=\sum_k(p_k^c+p_k^p)+P_c\); for Scheme B, \(R_{\text{total}}^{(t)}=\sum_m(r_1^m+r_2^m)\) and \(P_{\text{total}}^{(t)}=\sum_m(p_{11}^m+p_{12}^m+p_2^m)+P_c\).

\subsection{Overall Algorithm and Performance Analysis}
\subsubsection{Complete Three-Layer Framework}\label{sec:complete_framework}
Integrating the inner, middle, and outer layers constructs the comprehensive optimization framework outlined in Algorithm~\ref{alg:three_layer}. The process iterates through phase alignment, power allocation, and efficiency evaluation, culminating in final signal recovery.
\vspace{0.5em}
\hrule
\vspace{0.5em}
\begin{algorithm}\label{alg:three_layer}
    Three-Layer Decomposition Framework for RAQR-RSMA
    \vspace{0.5em}
    \hrule
    \vspace{0.5em}
    \begin{enumerate}
        \item \textbf{Input}: Complex channel $\mathbf{H}$, amplitude observations $\mathbf{z}$, LO field $\mathbf{b}$, bandwidth $B$, noise variance $\sigma^2$, power bounds $\{P_k^{\max}\}$, rate requirements $\{R_k^{\min}\}$, scheme selection flag $\mathcal{S} \in \{A, B\}$, tolerances $\varepsilon_{\mathrm{in}}, \varepsilon_{\mathrm{mid}}, \varepsilon_{\mathrm{out}}$.

        \item \textbf{Output}: Phase parameters $\boldsymbol{\psi}^*$, resource allocation, maximum EE $\eta^*$, and detected symbols $\hat{\mathbf{s}}$.

        \item \textbf{Stage 1: Initialization}
        \begin{enumerate}
            \item Initialize Dinkelbach parameter $\eta^{(0)} = 0$, outer iteration index $t = 0$.
            \item Initialize power allocation $\mathbf{p}^{(0)}$.
        \end{enumerate}

        \item \textbf{Stage 2: Iterative Optimization}
        \begin{enumerate}
            \item \textbf{Repeat}:
            \begin{itemize}
                \item \textbf{Layer 1: Inner Layer (Phase Alignment via Extended EM-GS)}.
                \begin{enumerate}
                    \item Initialize $\boldsymbol{\psi}^{(0)} = \mathbf{0}$.
                    \item Obtain $\hat{\mathbf{s}}^{(0)}$ from the known pilot observations via spectral initialization or pilot-based least squares in~\eqref{eq:pilot_init}.
                    \item \textbf{For} $i = 0, 1, \ldots$ until convergence:
                    \begin{itemize}
                        \item \textbf{E-step}: Compute posterior phase mean $\boldsymbol{\mu}^{(i)}$ with
                        $\mu_m^{(i)} = \angle(\mathbf{a}_m^H \hat{\mathbf{s}}^{(i)} + b_m)$,
                        and concentration $\kappa_m^{(i)} = \frac{2z_m|\mathbf{a}_m^H \hat{\mathbf{s}}^{(i)} + b_m|}{\sigma^2}$;
                        form
                        $\tilde{\mathbf{z}}^{(i)} = \mathbf{z} \circ e^{j\boldsymbol{\mu}^{(i)}} \circ R(\boldsymbol{\kappa}^{(i)})$.
                        \item \textbf{M-step}: Update
                        $\hat{\mathbf{s}}^{(i+1)} = (\mathbf{A}^H\mathbf{A})^{-1}\mathbf{A}^H\bigl(\tilde{\mathbf{z}}^{(i)} - \mathbf{b}\bigr)$.
                        \item \textbf{Gradient step}: Update $\boldsymbol{\psi}^{(i+1)} = \boldsymbol{\psi}^{(i)} - \mu \nabla J(\boldsymbol{\psi}^{(i)})$.
                    \end{itemize}
                    \item Set $\boldsymbol{\psi}^* \gets \boldsymbol{\psi}^{(i+1)}$, $\bar{h}_{m,k} = \Re\{h_{m,k}e^{j\psi_k^*}\}$, and $\hat{\mathbf{s}} \gets \hat{\mathbf{s}}^{(i+1)}$.
                    \item Obtain the phase-aligned real-valued channel $\tilde{\mathbf{H}}$ with entries $\bar{h}_{m,k} = \Re\{h_{m,k}e^{j\psi_k^*}\}$.
                \end{enumerate}

                \item \textbf{Layer 2: Middle Layer (Resource Allocation)}.
                \begin{enumerate}
                    \item Compute channel gains: $G_k = \sum_m |\tilde{H}_{m,k}|^2$, and sort $G_1 \le G_2 \le \cdots \le G_K$.
                    \item Execute Algorithm~\ref{alg:middle_layer} with scheme flag $\mathcal{S}$ and Dinkelbach parameter $\eta^{(t)}$ to obtain the resource allocation:
                    \begin{itemize}
                        \item \textbf{If} $\mathcal{S} = A$: obtain $\{p_k^{c*}, p_k^{p*}\}$.
                        \item \textbf{Else If} $\mathcal{S} = B$: obtain $\{p_{11}^{m*}, p_{12}^{m*}, p_2^{m*}, \alpha_m^*\}$.
                    \end{itemize}
                \end{enumerate}

                \item \textbf{Layer 3: Outer Layer (EE Update via Dinkelbach Iteration)}.
                \begin{enumerate}
                    \item \textbf{If} $\mathcal{S} = A$:
                     $R_{\text{total}}^{(t)} = \sum_{k=1}^{K}(r_k^c + r_k^p), \quad
                     P_{\text{total}}^{(t)} = \sum_{k=1}^{K}(p_k^c + p_k^p) + P_c.$

                    \item \textbf{Else If} $\mathcal{S} = B$: $R_{\text{total}}^{(t)} = \sum_{m=1}^{M}(r_1^m + r_2^m), \quad
                        P_{\text{total}}^{(t)} = \sum_{m=1}^{M}(p_{11}^m + p_{12}^m + p_2^m) + P_c.$
                    \item Compute objective
                    $F(\eta^{(t)}) = \bigl|R_{\text{total}}^{(t)} - \eta^{(t)} P_{\text{total}}^{(t)}\bigr|.$
                    \item \textbf{If} $F(\eta^{(t)}) < \varepsilon_{\mathrm{out}}$: \textbf{Break}.
                    \item Update
                    $\eta^{(t+1)} = \frac{R_{\text{total}}^{(t)}}{P_{\text{total}}^{(t)}},$
                    and increment $t \gets t+1$.
                \end{enumerate}
            \end{itemize}
            \item \textbf{Until} convergence criteria are met.
        \end{enumerate}

        \item \textbf{Stage 3: Signal Recovery}
        \begin{enumerate}
            \item Apply MMSE detection using the phase-aligned real channel:
            $\hat{\mathbf{s}} = \bigl(\tilde{\mathbf{H}}^T\tilde{\mathbf{H}} + \sigma^2 \mathbf{P}^{-1}\bigr)^{-1}\tilde{\mathbf{H}}^T\bigl(\tilde{\mathbf{z}} - \bar{\mathbf{b}}\bigr),$
            where $\mathbf{P} = \mathrm{diag}(p_1, \ldots, p_K)$ is the power allocation matrix, and $\bar{\mathbf{b}}$ is the equivalent real-valued LO component.
        \end{enumerate}

        \item \textbf{Return}: $\boldsymbol{\psi}^*$, resource allocation, $\eta^*$, and $\hat{\mathbf{s}}$.
    \end{enumerate}
    \vspace{0.5em}
    \hrule
    \vspace{0.5em}
\end{algorithm}

\subsubsection{Convergence Analysis}\label{sec:convergence}

The convergence of the proposed framework to a stationary point is analyzed layer by layer.

\begin{itemize}
    \item \textbf{Inner Layer Convergence:} The extended EM-GS algorithm produces a non-decreasing sequence of the expected complete-data log-likelihood and strictly reduces the residual imaginary energy. Hence, the phase alignment subproblem converges to a stationary point that yields a valid real-valued channel representation.
    \item \textbf{Middle Layer Convergence:} In symmetric deployments, the strict unimodality of $\mathrm{EE}(R)$ ensures that the golden-section search converges globally to the unique maximum $R^*$ at a linear rate. In asymmetric cases, for a given Dinkelbach parameter $\eta$, the BCD-PA algorithm produces a monotonically non-decreasing objective sequence that converges to a coordinate-wise stationary point of the subproblem.
    \item \textbf{Outer Layer Convergence:} Under Dinkelbach's fractional programming theory, the auxiliary function $F(\eta)$ is convex and strictly decreasing in $\eta$. The update $\eta^{(t+1)} = R(\mathbf{p}^{(t)})/P(\mathbf{p}^{(t)})$ strictly increases whenever $F(\eta^{(t)}) > 0$. Since the sum rate is bounded above and the transmit power is strictly positive, $\{\eta^{(t)}\}$ converges monotonically to the optimum $\eta^*$.
\end{itemize}

Since each layer monotonically improves its objective under bounded system constraints, the overall algorithm converges to a stationary point of the optimization problem.

\subsubsection{Complexity Comparison}

For Scheme A (Full RSMA), the BCD-PA algorithm updates $2K$ power variables per Dinkelbach iteration. Each coordinate update computes localized interference terms and applies water-filling projections, giving an inner-iteration complexity of $\mathcal{O}(K \cdot I_{\text{BCD}})$, where $I_{\text{BCD}}$ is the BCD loop count.

Scheme B (Paired RSMA) evaluates closed-form expressions over $M = \lfloor K/2 \rfloor$ independent pairs, with a bisection search of precision $\varepsilon_{\text{pair}}$. Its complexity is $\mathcal{O}(M \cdot \log_2(1/\varepsilon_{\text{pair}}))$, which is considerably lower than that of Scheme A for large $K$, since it avoids coupled iterative optimization within each pair.

For the common components, the inner EM-GS layer requires $\mathcal{O}(N_{\text{atom}} \cdot I_{\text{EM}})$ operations in both schemes, where $N_{\text{atom}}$ is the number of Rydberg vapor cells and $I_{\text{EM}}$ is the number of EM iterations. The outer Dinkelbach layer has the standard convergence complexity of $\mathcal{O}(\log(1/\varepsilon_{\text{out}}))$.

\section{Simulation Results and Analysis}
\label{sec:simulation}

In this section, we present numerical results to evaluate the convergence behavior and performance of the proposed energy-efficient uplink RSMA framework enabled by the RAQR.

Unless otherwise specified, the following common simulation parameters are adopted. We consider a single-cell uplink RSMA network, where one base station equipped with a $36$-element RAQR array serves three users. The effective noise power $N_0$ is determined by the quantum projection noise of the atomic ensemble. By substituting $N_a\approx 5\times 10^5$, $\tau_{\mathrm{coh}}\approx 10~\mu$s, and the typical dipole moment corresponding to the $52D_{5/2}\to 53P_{3/2}$ transition, we obtain $N_0\approx -114$ dBm. The circuit power consumption is set to $24.6$ W, and the maximum transmit power is $23$ dBm. The wireless channel includes both path loss and Rayleigh small-scale fading, and the effective channel gain is obtained by coherently combining the responses of the $36$ atomic array elements. The transmit power budget is swept from $-10$ to $30$ dBm with a step size of $2$ dB.

\begin{figure}[!htbp]
    \centering
    \includegraphics[width=0.9\linewidth]{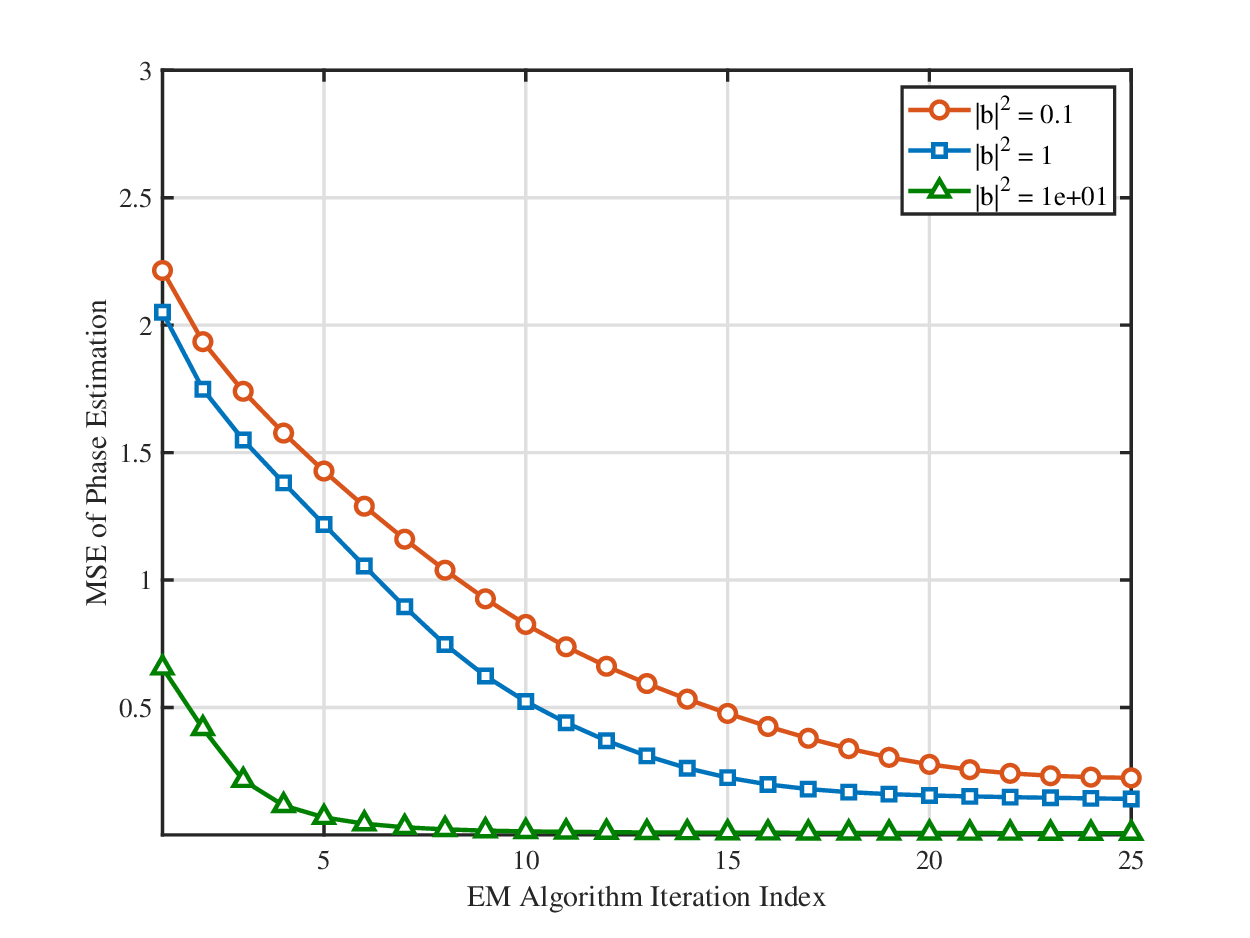}
    \caption{MSE of phase estimation versus EM algorithm iterations under different LO field intensities.}
    \label{fig:inner_convergence}
\end{figure}

Fig.~\ref{fig:inner_convergence} depicts the mean square error (MSE) of phase estimation versus the iteration index of the expectation-maximization (EM) algorithm under different local oscillator (LO) field intensities, denoted by $|b|^2$. It can be observed that the EM algorithm converges monotonically for all tested LO settings. Moreover, the LO field intensity has a clear impact on both the convergence speed and the final estimation accuracy. When the LO field is strong, e.g., $|b|^2=10$, the MSE decreases rapidly and stabilizes within approximately $5$ to $8$ iterations. In contrast, when the LO field is weak, e.g., $|b|^2=0.1$, the algorithm requires more than $20$ iterations to converge and settles at a higher steady-state MSE. This behavior indicates that a sufficiently strong LO field can effectively reduce the phase ambiguity in amplitude-only RAQR measurements, thereby improving the robustness of the EM-based phase retrieval process.

\begin{figure}[!htbp]
    \centering
    \includegraphics[width=0.9\linewidth]{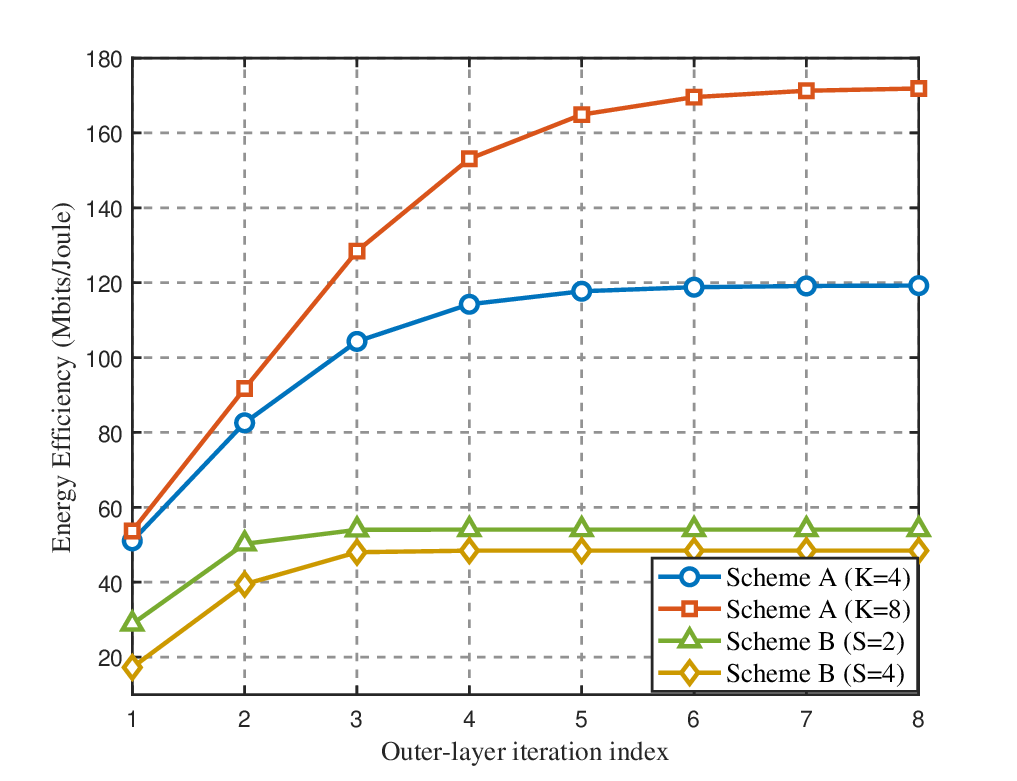}
    \caption{System EE versus Dinkelbach iterations for Scheme A and Scheme B.}
    \label{fig:outer_convergence}
\end{figure}

The convergence of the outer-layer optimization is shown in Fig.~\ref{fig:outer_convergence}, where the system EE is plotted against the Dinkelbach iteration index for Scheme~A (full RSMA in the asymmetric-rate setting solved by BCD-PA) and Scheme~B (paired RSMA) under different user scales. In all considered cases, the system EE increases monotonically and converges to a stable value within a small number of iterations, which verifies the effectiveness of the proposed fractional programming framework. For full RSMA, the number of required iterations increases with the total number of users $K$. For example, when $K=8$, convergence is typically reached after about $8$ iterations. At the same time, full RSMA achieves a higher EE because it can more effectively exploit multiuser diversity. By comparison, paired RSMA converges faster, typically within $2$ to $3$ iterations, but its final EE is lower than that of full RSMA. In addition, for paired RSMA, increasing the number of user pairs $S$ from $2$ to $4$ further reduces the achievable EE, since orthogonal bandwidth partitioning among more pairs limits the available spectral resources for each pair.

\begin{figure}[!htbp]
    \centering
    \includegraphics[width=0.9\linewidth]{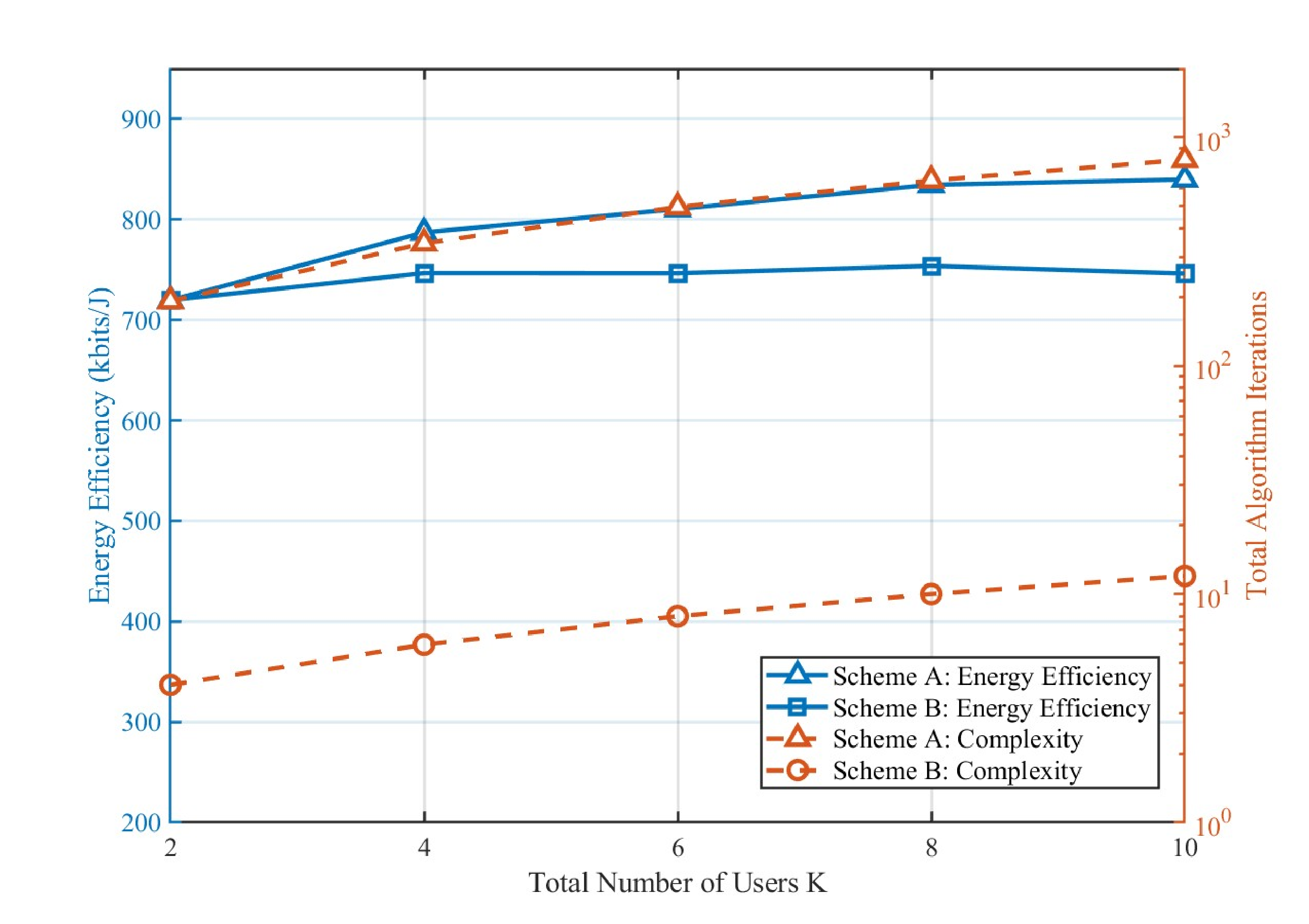}
    \caption{Trade-off between EE and computational complexity versus the total number of users $K$ under asymmetric rate requirements.}
    \label{fig:complexity_tradeoff}
\end{figure}

Fig.~\ref{fig:complexity_tradeoff} further investigates the trade-off between EE and computational complexity as the total number of users $K$ increases from $2$ to $10$ under asymmetric rate requirements. The left vertical axis corresponds to EE, while the right vertical axis indicates the total number of algorithm iterations on a logarithmic scale. As $K$ increases, the EE achieved by Scheme~A gradually improves, which can be attributed to the stronger multiuser diversity available in larger systems. However, this gain is accompanied by a rapid increase in computational cost. In particular, the total number of iterations approaches $10^3$ when $K=10$. In contrast, Scheme~B decomposes the original problem into multiple lower-dimensional intra-pair optimizations, which substantially reduces the computational burden. Even when $K=10$, Scheme~B requires fewer than $20$ iterations while maintaining a relatively stable EE level. These results indicate that Scheme~B provides a practical low-complexity alternative, with a moderate loss in EE performance compared with Scheme~A in large-scale multi-user settings.

\begin{figure}[!htbp]
\centering
\includegraphics[width=0.9\linewidth]{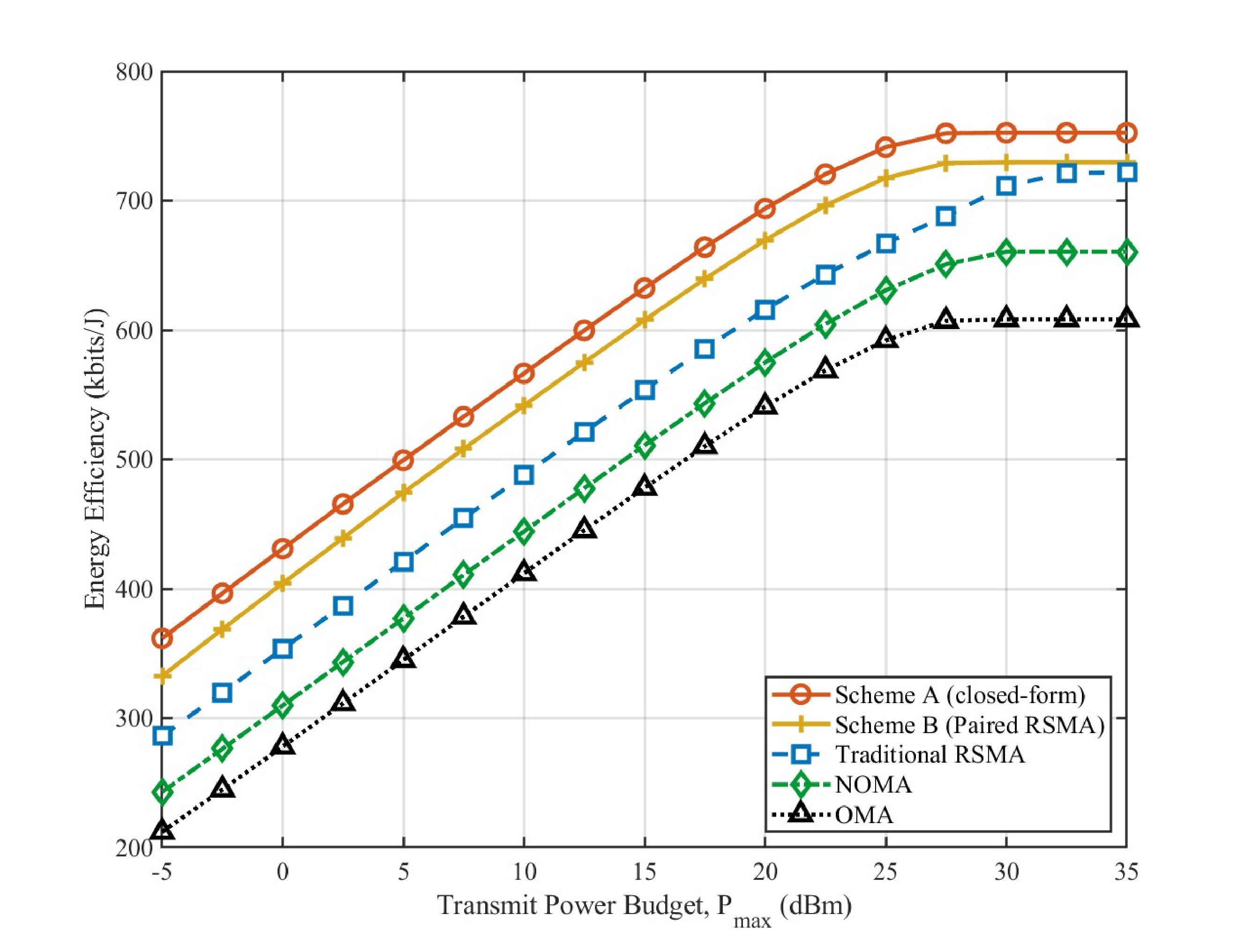}
\caption{EE versus transmit power budget under symmetric rate requirements.}
\label{fig:symmetric_ee}
\end{figure}

Fig.~\ref{fig:symmetric_ee} shows the EE as a function of the transmit power budget under symmetric rate requirements. As the power budget increases from $-5$ dBm to $35$ dBm, the EE of all considered schemes first increases and then gradually saturates in the high-power region, particularly beyond $25$ dBm. This trend indicates that, under the EE maximization criterion, increasing the transmit power beyond the optimal operating region does not provide additional EE improvement. Among all considered schemes, the proposed Scheme~A with the closed-form symmetric solution achieves the highest EE over the entire power range, due to its ability to fully exploit the available multiuser diversity. Scheme~B, i.e., paired RSMA, achieves the second-best performance and remains close to Scheme~A. Both proposed schemes outperform the benchmark methods including traditional RSMA, NOMA, and OMA. In particular, OMA exhibits the lowest EE because orthogonal resource allocation is less efficient in multi-user interference-limited scenarios.

\begin{figure}[!htbp]
\centering
\includegraphics[width=0.9\linewidth]{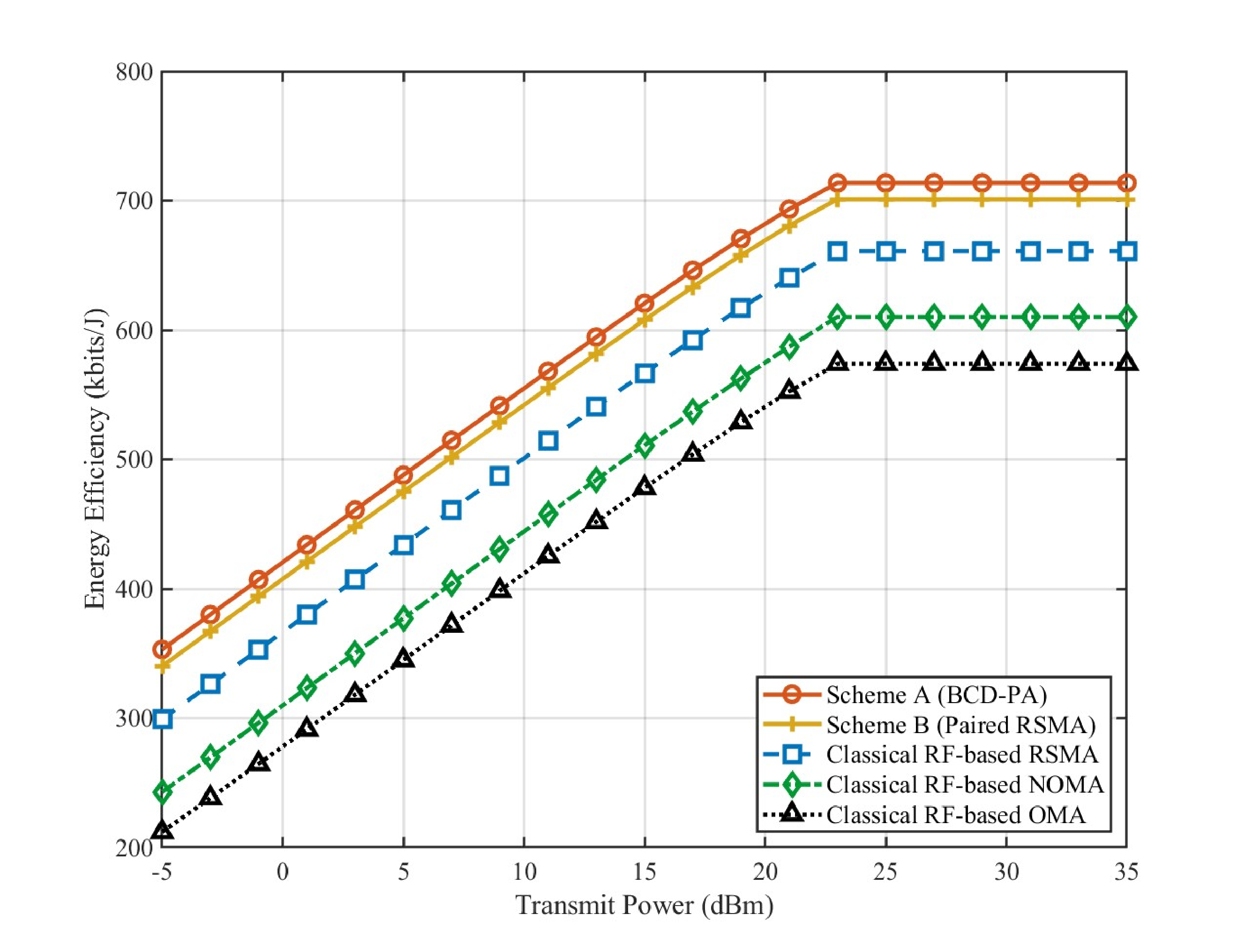}
\caption{EE versus transmit power budget under asymmetric rate requirements.}
\label{fig:asymmetric_ee}
\end{figure}

Fig.~\ref{fig:asymmetric_ee} presents the EE performance under asymmetric rate requirements. The overall trend is similar to that observed in the symmetric case. Specifically, the EE increases in the low-power region and then approaches a plateau when the power budget exceeds approximately $20$ dBm. Under heterogeneous channel and rate conditions, Scheme~A dynamically adjusts the power allocation between common and private streams through the BCD-PA algorithm, and therefore achieves the best EE among all compared schemes. Scheme~B maintains a slightly lower but still competitive EE, while avoiding the higher optimization complexity associated with Scheme~A. In addition, both RAQR-enabled schemes consistently outperform the classical RF-based RSMA, NOMA, and OMA benchmarks. Classical NOMA shows a more noticeable performance degradation in this asymmetric setting, mainly because its fixed SIC decoding order is less adaptable to heterogeneous user conditions.

\begin{figure}[!htbp]
\centering
\includegraphics[width=0.9\linewidth]{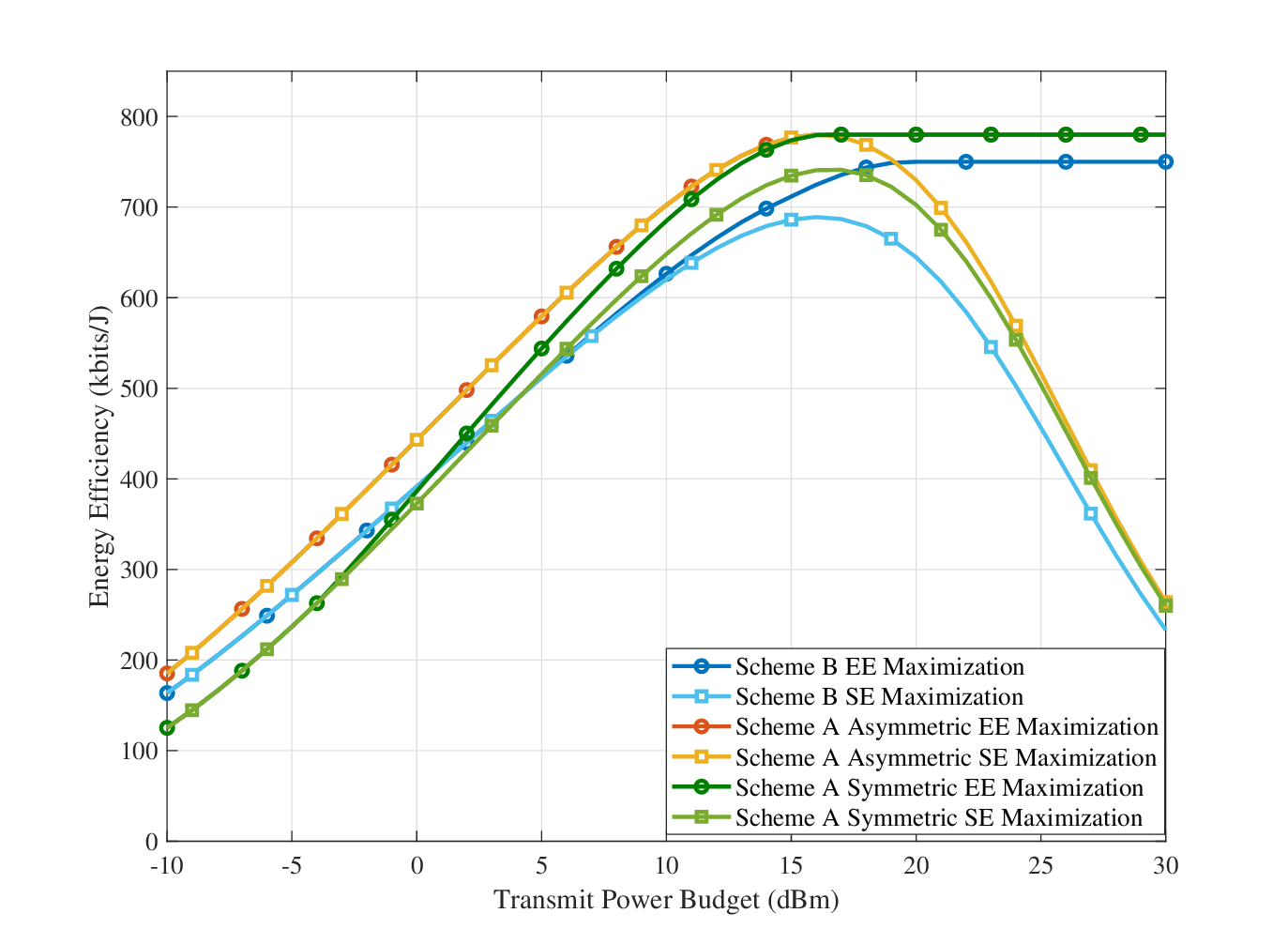}
\caption{Performance comparison under three channel conditions.}
\label{fig:ee_vs_se}
\end{figure}

To further evaluate the proposed energy-efficient design, Fig.~\ref{fig:ee_vs_se} compares the EE achieved under EE maximization and spectral efficiency (SE) maximization for six configurations over the transmit power budget range from $-10$ to $30$ dBm. It can be observed that, in all three scenarios, the SE-maximizing strategies tend to use the entire available transmit power budget, and their EE decreases once the power budget exceeds the optimal operating point. In contrast, the EE-maximizing schemes maintain a stable high-power plateau by avoiding unnecessary power expenditure after the optimal EE has been reached. Fig.~\ref{fig:ee_vs_se} also provides a direct comparison among the proposed schemes. The EE-maximizing Scheme~A achieves the highest EE, reaching approximately $780$ to $800$ Kbits/J, while Scheme~B optimized for EE achieves around $700$ Kbits/J. In addition, the symmetric and asymmetric versions of Scheme~A perform closely in the high-power regime, indicating that the asymmetric BCD-PA algorithm can effectively approach the performance of the KKT-based closed-form solution for the symmetric case. 

\section{Conclusions}
\label{section:Conclusion}
This paper investigated an RAQR-empowered uplink RSMA network for energy efficiency (EE) maximization. To address the amplitude-only measurement constraint of RAQRs, a phase alignment mechanism was introduced to obtain a tractable linearized channel model. The resulting non-convex fractional EE problem was handled via Dinkelbach's method. For symmetric IoT scenarios, exact closed-form power allocation was derived from the KKT conditions, enabling a globally optimal bisection search with low complexity. For asymmetric settings, a generalized BCD-PA framework was developed for heterogeneous rate requirements. In addition, a paired resource allocation scheme (Scheme B) based on Lambert-W-function closed-form solutions was proposed to further reduce complexity. Numerical results showed that Scheme A achieves the highest EE, while Scheme B provides a favorable complexity-performance trade-off, making it suitable for massive IoT deployment in RSMA-based 6G networks. Phase-sensitive RAQR readout remains a topic for future work due to its additional hardware requirements.


\balance
 \bibliographystyle{ieeetr}
 \bibliography{my_references}

@article{mao2022rate,
  title={Rate-splitting multiple access: Fundamentals, survey, and future research trends},
  author={Mao, Yijie and Dizdar, Onur and Clerckx, Bruno and Schober, Robert and Popovski, Petar and Poor, H Vincent},
  journal={IEEE Commun. Surveys Tuts.},
  volume={24},
  number={4},
  pages={2073--2126},
  year={2022},
  publisher={IEEE}
}

@article{clerckx2021noma,
  title={Is NOMA efficient in multi-antenna networks? A critical look at next generation multiple access techniques},
  author={Clerckx, Bruno and Mao, Yijie and Schober, Robert and Jorswieck, Eduard and Love, David J and Yuan, Jinhong and Hanzo, Lajos and Li, Geoffrey Ye and Larsson, Erik G and Caire, Giuseppe},
  journal={arXiv preprint arXiv:2101.04802},
  year={2021}
}

@inproceedings{zhu2017rate,
  title={A rate-splitting non-orthogonal multiple access scheme for uplink transmission},
  author={Zhu, Ye and Wang, Xianbin and Zhang, Zhaoyang and Chen, Xiaoming and Chen, Yan},
  booktitle={2017 9th International Conference on Wireless Communications and Signal Processing (WCSP)},
  pages={1--6},
  year={2017},
  organization={IEEE}
}

@article{yang2020sum,
  title={Sum-rate maximization of uplink rate splitting multiple access (RSMA) communication},
  author={Yang, Zhaohui and Chen, Mingzhe and Saad, Walid and Xu, Wei and Shikh-Bahaei, Mohammad},
  journal={IEEE Trans. Mobile Comput.},
  volume={21},
  number={7},
  pages={2596--2609},
  year={2020},
  publisher={IEEE}
}

@article{sun2023joint,
  title={Joint receive and passive beamforming optimization for RIS-assisted uplink RSMA systems},
  author={Sun, Qiang and Liu, Hongwu and Yan, Shen and Tsiftsis, Theodoros A and Yuan, Jinhong},
  journal={IEEE Wireless Commun. Lett.},
  volume={12},
  number={7},
  pages={1204--1208},
  year={2023},
  publisher={IEEE}
}

@article{liu2023downlink,
  title={Downlink energy efficiency maximization for RSMA-UAV assisted communications},
  author={Liu, Xin and Feng, Jihan and Li, Feng and Leung, Victor CM},
  journal={IEEE Wireless Commun. Lett.},
  volume={13},
  number={1},
  pages={98--102},
  year={2023},
  publisher={IEEE}
}

@ARTICLE{10643592,
  author={Can, Mehmet and Altunbas, Ibrahim},
  journal={IEEE Wireless Commun. Lett.},
  title={Channel-Dependent Power Allocation and Decoding Order for RSMA-Based Uplink Satellite Networks},
  year={2024},
  volume={13},
  number={10},
  pages={2822-2826},
  doi={10.1109/LWC.2024.3448297}
}

@article{can2025data,
  title={Data-oriented uplink RSMA systems: Performance analysis and design insight},
  author={Can, Mehmet and Ilter, Mehmet C and Altunbas, Ibrahim and Valkama, Mikko},
  journal={IEEE Trans. Commun.},
  volume={73},
  number={8},
  pages={6639--6654},
  year={2025},
  publisher={IEEE}
}

@article{liu2025joint,
  title={Joint optimization of communication latency and platoon control based on uplink RSMA for future V2X networks},
  author={Liu, Gang and Hu, Jiewen and Ma, Zheng and Fan, Pingzhi and Yu, F Richard},
  journal={IEEE Trans. Veh. Technol.},
  year={2025},
  publisher={IEEE}
}

@ARTICLE{11052254,
  author={Nyoman Apraz Ramatryana, I and Bayu Satrya, Gandeva and Karna, Nyoman Bogi Aditya and Adi Paramartha Putra, Made},
  journal={IEEE Commun. Lett.},
  title={Uplink RSMA-Assisted Slotted ALOHA With Adaptive Traffic Load for Massive IoT},
  year={2025},
  volume={29},
  number={9},
  pages={2013-2017},
  doi={10.1109/LCOMM.2025.3583590}
}

@INPROCEEDINGS{11161157,
  author={Tegos, Apostolos A. and Xiao, Yue and Tegos, Sotiris A. and Karagiannidis, George K. and Diamantoulakis, Panagiotis D.},
  booktitle={ICC 2025 - IEEE International Conference on Communications},
  title={Extending the Capacity Region with Distributed Uplink Rate Splitting Multiple Access (DU-RSMA)},
  year={2025},
  volume={},
  number={},
  pages={2863-2868}
}

@ARTICLE{9852986,
  author={Abbasi, Omid and Yanikomeroglu, Halim},
  journal={IEEE Trans. Wireless Commun.},
  title={Transmission Scheme, Detection and Power Allocation for Uplink User Cooperation With NOMA and RSMA},
  year={2023},
  volume={22},
  number={1},
  pages={471-485},
  doi={10.1109/TWC.2022.3195532}
}

@ARTICLE{10330667,
  author={Hu, Jiewen and Liu, Gang and Ma, Zheng and Xiao, Ming and Fan, Pingzhi},
  journal={IEEE Wireless Commun. Lett.},
  title={Low-Complexity Resource Allocation for Uplink RSMA in Future 6G Wireless Networks},
  year={2024},
  volume={13},
  number={2},
  pages={565-569},
  doi={10.1109/LWC.2023.3336949}
}

@article{sadeghabadi2025achievable,
  title={Achievable Splitting and Decoding in the Multiple Access Channel: A Hierarchical Approach},
  author={Sadeghabadi, Elaheh and Blostein, Steven},
  journal={IEEE Trans. Wireless Commun.},
  year={2025},
  publisher={IEEE}
}

@article{ding2015impact,
  title={Impact of user pairing on 5G nonorthogonal multiple-access downlink transmissions},
  author={Ding, Zhiguo and Fan, Pingzhi and Poor, H Vincent},
  journal={IEEE Trans. Veh. Technol.},
  volume={65},
  number={8},
  pages={6010--6023},
  year={2015},
  publisher={IEEE}
}

@article{degen2017quantum,
  title={Quantum Sensing},
  author={Degen, Christian L and Reinhard, Friedemann and Cappellaro, Paola},
  journal={Rev. Mod. Phys.},
  volume={89},
  number={3},
  pages={035002},
  year={2017},
  publisher={APS}
}

@article{gong2025rydberg,
  title={Rydberg Atomic Quantum Receivers for Classical Wireless Communication and Sensing},
  author={Gong, Tierui and Chandra, Aveek and Yuen, Chau and Guan, Yong Liang and Dumke, Rainer and See, Chong Meng Samson and Debbah, Merouane and Hanzo, Lajos},
  journal={IEEE Wireless Commun.},
  year={2025},
  publisher={IEEE}
}

@article{chen2025harnessing,
  title={Harnessing Rydberg Atomic Receivers: From Quantum Physics to Wireless Communications},
  author={Chen, Yuanbin and Guo, Xufeng and Yuen, Chau and Zhao, Yufei and Guan, Yong Liang and See, Chong Meng Samson and D{\'e}bbah, Merouane and Hanzo, Lajos},
  journal={arXiv preprint arXiv:2501.11842},
  year={2025}
}

@article{saffman2010quantum,
  title={Quantum Information with Rydberg Atoms},
  author={Saffman, Mark and Walker, Thad G and M{\o}lmer, Klaus},
  journal={Rev. Mod. Phys.},
  volume={82},
  number={3},
  pages={2313--2363},
  year={2010},
  publisher={APS}
}

@article{anderson2021atomic,
  author={Anderson, David Alexander and Sapiro, Rachel Elizabeth and Raithel, Georg},
  journal={IEEE Trans. Antennas Propag.},
  title={An Atomic Receiver for AM and FM Radio Communication},
  year={2021},
  volume={69},
  number={5},
  pages={2455-2462},
  doi={10.1109/TAP.2020.2987112}
}

@article{artusio2022modern,
  title={Modern RF Measurements with Hot Atoms: A Technology Review of Rydberg Atom-based Radio Frequency Field Sensors},
  author={Artusio-Glimpse, Alexandra and Simons, Matthew T and Prajapati, Nikunjkumar and Holloway, Christopher L},
  journal={IEEE Microw. Mag.},
  volume={23},
  number={5},
  pages={44--56},
  year={2022},
  publisher={IEEE}
}

@article{wang2022quantum,
  title={Quantum-enabled 6G Wireless Networks: Opportunities and Challenges},
  author={Wang, Chonggang and Rahman, Akbar},
  journal={IEEE Wireless Commun.},
  volume={29},
  number={1},
  pages={58--69},
  year={2022},
  publisher={IEEE}
}

@article{cui2025rydberg,
  title={Rydberg Atomic Receiver: Next Frontier of Wireless Communications},
  author={Cui, Mingyao and Zeng, Qunsong and Huang, Kaibin},
  journal={IEEE Commun. Mag.},
  year={2025},
  publisher={IEEE}
}

@article{fan2015atom,
  title={Atom Based RF Electric Field Sensing},
  author={Fan, Haoquan and Kumar, Santosh and Sedlacek, Jonathan and K{\"u}bler, Harald and Karimkashi, Shervin and Shaffer, James P.},
  journal={J. Phys. B: At. Mol. Opt. Phys.},
  volume={48},
  number={20},
  pages={202001},
  year={2015},
  doi={10.1088/0953-4075/48/20/202001}
}

@article{meyer2018digital,
  title={Digital Communication with Rydberg Atoms and Amplitude-modulated Microwave Fields},
  author={Meyer, David H and Cox, Kevin C and Fatemi, Fredrik K and Kunz, Paul D},
  journal={Appl. Phys. Lett.},
  volume={112},
  number={21},
  year={2018},
  publisher={AIP Publishing}
}

@article{yuan2023rydberg,
  title={A Rydberg Atom-based Receiver with Amplitude Modulation Technique for the Fifth-generation Millimeter-wave Wireless Communication},
  author={Yuan, Jinpeng and Jin, Ting and Xiao, Liantuan and Jia, Suotang and Wang, Lirong},
  journal={IEEE Antennas Wireless Propag. Lett.},
  volume={22},
  number={10},
  pages={2580--2584},
  year={2023},
  publisher={IEEE}
}

@article{jing2020atomic,
  title={Atomic Superheterodyne Receiver Based on Microwave-dressed Rydberg Spectroscopy},
  author={Jing, Mingyong and Hu, Ying and Ma, Jie and Zhang, Hao and Zhang, Linjie and Xiao, Liantuan and Jia, Suotang},
  journal={Nat. Phys.},
  volume={16},
  number={9},
  pages={911--915},
  year={2020},
  publisher={Nature Publishing Group UK London}
}

@article{simons2021rydberg,
  title={Rydberg Atom-based Sensors for Radio-frequency Electric Field Metrology, Sensing, and Communications},
  author={Simons, Matthew T and Artusio-Glimpse, Alexandra B and Robinson, Amy K and Prajapati, Nikunjkumar and Holloway, Christopher L},
  journal={Meas. Sensors},
  volume={18},
  pages={100273},
  year={2021},
  publisher={Elsevier}
}

@article{fancher2021rydberg,
  title={Rydberg Atom Electric Field Sensors for Communications and Sensing},
  author={Fancher, Charles T and Scherer, David R and John, Marc C St and Marlow, Bonnie L Schmittberger},
  journal={IEEE Trans. Quantum Eng.},
  volume={2},
  pages={1--13},
  year={2021},
  publisher={IEEE}
}

@article{yuan2023quantum,
  title={Quantum Sensing of Microwave Electric Fields Based on Rydberg Atoms},
  author={Yuan, Jian and Yang, Wenguo and Jing, Mingyong and Zhang, Hao and Jiao, Yuechun and Li, Wei and Zhang, Lin and Xiao, Liantuan and Jia, Suotang},
  journal={Rep. Prog. Phys.},
  volume={86},
  number={10},
  pages={106001},
  year={2023},
  doi={10.1088/1361-6633/acf22f}
}

@article{meyer2020assessment,
  title={Assessment of Rydberg Atoms for Wideband Electric Field Sensing},
  author={Meyer, David H and Castillo, Zachary A and Cox, Kevin C and Kunz, Paul D},
  journal={J. Phys. B: At. Mol. Opt. Phys.},
  volume={53},
  number={3},
  pages={034001},
  year={2020},
  publisher={IOP Publishing}
}

@article{santamaria2022comparison,
  title={Comparison of Noise Temperature of Rydberg-atom and Electronic Microwave Receivers},
  author={Santamaria-Botello, Gabriel and Verploegh, Shane and Bottomley, Eric and Popovic, Zoya},
  journal={arXiv preprint arXiv:2209.00908},
  year={2022}
}

@article{simons2019rydberg,
  title={A Rydberg Atom-based Mixer: Measuring the Phase of a Radio Frequency Wave},
  author={Simons, Matthew T and Haddab, Abdulaziz H and Gordon, Joshua A and Holloway, Christopher L},
  journal={Appl. Phys. Lett.},
  volume={114},
  number={11},
  year={2019},
  publisher={AIP Publishing}
}

@article{cui2025towards,
  title={Towards Atomic MIMO Receivers},
  author={Cui, Mingyao and Zeng, Qunsong and Huang, Kaibin},
  journal={IEEE J. Sel. Areas Commun.},
  year={2025},
  publisher={IEEE}
}

@inproceedings{gong2025equivalent,
  title={An Equivalent Baseband Signal Model for Rydberg Atomic Quantum Receiver Aided Wireless Communications and Sensing},
  author={Gong, Tierui and Sun, Jiaming and Yuen, Chau and Hu, Guangwei and Zhao, Yufei and Guan, Yong Liang and See, Chong Meng Samson and Debbah, M{\'e}rouane and Hanzo, Lajos},
  booktitle={2025 International Conference on Quantum Communications, Networking, and Computing (QCNC)},
  pages={200--204},
  year={2025},
  organization={IEEE}
}

@article{liu2025sa,
  title={SA-MIMO: Scalable Quantum-Based Wireless Communications},
  author={Liu, Jiuyu and Ma, Yi and Tafazolli, Rahim},
  journal={arXiv preprint arXiv:2504.19170},
  year={2025}
}

@article{wang2020tight,
  title={Tight Capacity Bounds for Indoor Visible Light Communications with Signal-dependent Noise},
  author={Wang, Jin-Yuan and Fu, Xian-Tao and Lu, Rong-Rong and Wang, Jun-Bo and Lin, Min and Cheng, Julian},
  journal={IEEE Trans. Wireless Commun.},
  volume={20},
  number={3},
  pages={1700--1713},
  year={2020},
  publisher={IEEE}
}

@inproceedings{cui2024mimo,
  author={Cui, Mingyao and Zeng, Qunsong and Huang, Kaibin},
  booktitle={ICC 2025 - IEEE International Conference on Communications},
  title={MIMO Precoding for Rydberg Atomic Receivers Enabled Wireless Communications},
  year={2025},
  pages={4792-4797},
  doi={10.1109/ICC52391.2025.11160871}
}

@article{xu2025channel,
  author={Xu, Bokai and Zhang, Jiayi and Chen, Zhongtao and Cheng, Bingyang and Liu, Ziheng and Wu, Yik-Chung and Ai, Bo},
  journal={IEEE Wireless Commun. Lett.},
  title={Channel Estimation for Rydberg Atomic Receivers},
  year={2025},
  volume={14},
  number={9},
  pages={2957-2961},
  doi={10.1109/LWC.2025.3584040}
}

@article{gong2025rydbergMIMO,
  title={Rydberg Atomic Quantum MIMO Receivers for the Multi-user Uplink},
  author={Gong, Tierui and Yuen, Chau and See, Chong Meng Samson and Debbah, M{\'e}rouane and Hanzo, Lajos},
  journal={arXiv preprint arXiv:2506.01355},
  year={2025}
}

@article{peng2025ris,
  title={RIS-assisted Atomic MIMO Receiver},
  author={Peng, Qihao and Liu, Jiuyu and Luo, Qu and Ma, Yi and Xiao, Pei and Elkashlan, Maged and Karagiannidis, George K},
  journal={arXiv preprint arXiv:2510.15763},
  year={2025}
}

@article{cui2025rydberg2,
  title={Rydberg Atomic Receivers for Multi-Band Communications and Sensing},
  author={Cui, Mingyao and Zeng, Qunsong and Wang, Zhanwei and Huang, Kaibin},
  journal={arXiv preprint arXiv:2505.24168},
  year={2025}
}

@article{wang2025multi,
  title={Multi-Carrier Rydberg Atomic Quantum Receivers with Enhanced Bandwidth Feature for Communication and Sensing},
  author={Wang, Huizhi and Gong, Tierui and Bj{\"o}rnson, Emil and Yuen, Chau},
  journal={arXiv preprint arXiv:2510.10473},
  year={2025}
}

@article{gong2025rydberg2,
  author={Gong, Tierui and Yuen, Chau and See, Chong Meng Samson and Debbah, M{\'e}rouane and Hanzo, Lajos},
  journal={IEEE Trans. Veh. Technol.},
  title={Rydberg Atomic Quantum Receivers for Multi-Target DOA Estimation},
  year={2025},
  pages={1-6},
  doi={10.1109/TVT.2025.3644954}
}

@article{chen2025polarization,
  title={Polarization-aware DOA detection relying on a single Rydberg atomic receiver},
  author={Chen, Yuanbin and Yuen, Chau and Arumugam, Darmindra and See, Chong Meng Samson and Debbah, M{\'e}rouane and Hanzo, Lajos},
  journal={arXiv preprint arXiv:2508.17179},
  year={2025}
}

\end{document}